\documentclass[twocolumn]{aa}
\usepackage[utf8]{inputenc}
\usepackage{natbib}
\usepackage{graphicx}
\usepackage{placeins}
\usepackage{color}
\usepackage{amsmath}
\usepackage{natbib}
\usepackage{arydshln}
\usepackage{tikz}
\usepackage{hyperref}
\usepackage{orcidlink}
\newcommand{\FIG}[1]{#1}
\usepackage{xcolor}
\usepackage{soul}

\title{Two-fluid reconnection jets in a gravitationally stratified atmosphere.}
\author{B. Popescu Braileanu
          \inst{1}
          \and
          R. Keppens \inst{1}\orcidlink{0000-0003-3544-2733}}
\institute{Centre for mathematical Plasma Astrophysics, KU Leuven, 3001 Leuven, Belgium,\\ \email{beatriceannemone.popescubraileanu@kuleuven.be} }
\date{}

\date{Received XXXX; Accepted XXXX}
 
\abstract{
{\it Context.} 
The density decreases exponentially with height in the solar gravitationally stratified atmosphere, therefore the 
collisional coupling between the ionized plasma and the neutrals also decreases.
Reconnection is a process observed at all heights in the solar atmosphere.
\\{\it Aims.}
{Here, we investigate} the role of collisions between ions and neutrals on the reconnection process occurring at various heights in the atmosphere.
\\{\it Methods.}
We perform simulations of magnetic reconnection induced by a localized resistivity in a gravitationally stratified atmosphere,
where we vary the height of the initial reconnection X-point. 
We compare a magnetohydrodynamic (MHD)  model and two two-fluid configurations:
one where the collisional coupling is calculated from local plasma parameters and another
where the coupling is decreased, so that collisional effects are enhanced. The latter setup has a more representative solar collisionality regime.
\\{\it Results.} 
Simulations in a stratified atmosphere show similar structures in MHD and two-fluid simulations with strong coupling.
However, when collisional effects are increased to attain representative parameter regimes, we find a nonlinear runaway instability, which separates the plasma-neutral densities across the current sheet (CS). With increased collisional effects, the initial decoupling in velocity heats the neutrals and this sets up a nonlinear feedback where neutrals migrate outside the CS, replacing charged particles which accumulate towards the center of the CS.
\\{\it Conclusions.}
The reconnection rate has a maximum value around 0.1, similar for both reconnection heights, and is consistent with the use of a localized enhanced resistivity used in all three models.
The initial stages of plasmoid formation, observed near the end of our simulations, is influenced
by the outflow from the primary reconnection point, rather than by collisions. We synthesize optically thin emission for both MHD and two-fluid models, which can show a very different evolution 
when the charged particle density is used instead of the total density. Our simulations have relevance for observed plasmoid features associated with chromospheric to low coronal flare events.
}

\begin{document}

\maketitle

\section{Introduction}
The temperature increases steeply towards the corona of the quiet sun, throughout the transition region located at $\approx$~2.5~Mm, and coronal plasma becomes fully ionized \citep{ionFr}.
Observations showed that field lines  below $\approx$~2.5~Mm can change their connectivity in about ~1.5~h,
suggesting fast reconnection mechanisms in the solar chromosphere \citep{recObs,recReview}. 

The reconnection mechanism involved in solar flares { releases a huge amount of magnetic energy}, therefore it is believed that the 
reconnection is driven at large scales, evolving towards a Petschek-type configuration \citep{sato}.
Petschek \citep{petschek} proposed a stationary 2D reconnection model that extends a Sweet-Parker \citep{sweet,parker} layer (the diffusion region) at the center with standing slow-mode shocks from its ends.
{ The Sweet-Parker reconnection rate scales as $\propto S^{-1/2}$, where the nondimensional Lundquist number, $S$ is defined as $S=v_A L/\eta$, with $v_A$ being the Alfv\'en 
speed, $L$ a hydrodynamic characteristic length and $\eta$ the value of the resistivity (in units of m$^2$/s)}
This classical resistive MHD model proposed by \cite{petschek} predicts reconnection rates much higher than the Sweet-Parker model, { where it 
scales as $\propto \text{ln}^{-1}(S)$. For a value of $S=10^6$, the Petschek-type reconnection gives a reconnection rate 65 times larger than the Sweet-Parker model \citep{slavaRec1}.} 
With larger resistivity, the slow shocks become wider \citep{sato}.
The magnetic energy decreases, being converted into thermal energy through Ohmic heating near the diffusion region, and to 
kinetic energy through the work done by the Lorentz force, away from the diffusion region, especially at the location of the slow shocks \citep{ugai}. In simulations of driven reconnection, anomalous resistivity is assumed in order to impede the unbounded growth of the current density \citep{sato,ugai,Uzdensky_2003}.

Anomalous resistivity can be produced by ion acoustic turbulence, and is then a function of the ion electron drift velocity, as shown by \cite{anomalousRes}. 
A more simplified, but frequently adopted model of anomalous resistivity is a space localized resistivity around the X-point, which is also used in this paper.
{ Unlike anomalous resistivity which depends on the current density, the localized resistivity only depends on space, but both prescriptions lead to fast magnetic reconnection \citep{biskamp-loc}.}
This locally enhanced resistivity leads to fast reconnection, however, it is not very clear  whether fast reconnection is due to a high value of the local resistivity in the diffusion region, or to the localization \citep{biskamp-loc}. Numerical 2D resistive MHD simulations found that having a flat local resistivity profile near the X-point can induce spontaneous symmetry breaking in the otherwise symmetric Petschek configuration \cite{2014Baty}. In this paper, we will have an { additional} broken up-down symmetry due to gravity from the beginning, and extend Petschek-reconnection findings to a two-fluid, plasma-neutral setting.

Many simulations study the standard solar flare scenario, where a vertical current sheet (CS) evolves to form post-flare loops.
\cite{Takasao2015} studied post-flare loops in the MHD approximation using a 2.5D setup, where gravity is neglected and thermal conductivity is adopted along the field lines. In their simulations, Petschek-type reconnection develops because of localized resistivity 
and the slow shocks are essentially isothermal due to effective thermal conductivity.
\cite{2021wang} study reconnection in a 2.5D MHD  setup, using a spatially localized resistivity and an analytic density profile, with uniform pressure. In their model, gravity is neglected, while anisotropic thermal conductivity is incorporated. 
The reconnection rate was found to be slightly smaller when plasma $\beta$ increases, and the rate is also smaller with thermal conductivity. The reconnection rate reaches a maximal value of 0.01.
The authors show that reconnection at the termination shock due to interaction between magnetic islands formed along the primary current sheet (CS) and the magnetic arcade below is almost as important as reconnection in the main CS for releasing magnetic energy. Jets produced by MHD simulations of Petschek reconnection in a 2D setup without gravity have properties of small-scale flares observed in the solar atmosphere \citep{1999gabor}. State-of-the-art solar flare simulations extend these efforts with the inclusion of gravitational stratification, and even
include the effect of fast electron beams that self-consistently interact with large scale MHD simulations, identifying many ingredients found in actual observations \citep{wenzhi2}. The step to full 3D MHD simulations, including gravity, thermal conduction and reproducing turbulent regions consistent with observed non-thermal velocities was made in \cite{wenzhi}. EUV synthetic images produced from these 3D flare models show very good agreement with observations. However, extension to plasma-neutral setups are needed to address more chromospheric flare counterparts, and this work is a first step towards that goal.

Reconnection jets have been observed at all heights in the solar atmosphere, from photosphere to corona \citep{obsJets2} in both cool and hot spectral lines.
 \cite{2013takasao} suggest that spicules are  chromospheric jets in emerging flux regions, which disappear in chromospheric line images before returning, probably due to heating.
Hot and cool jets observed by \cite{obsJets} were suggested to form in the lower corona or upper chromosphere. Aspects of coronal X-ray jets were successfully reproduced in simulations by \cite{1996yoko}.
Chromospheric anemone jets with  velocities of 10~km/s comparable to the local Alfv\'en speed were observed in the upper chromosphere, but could not be  observed in the lower chromosphere, where  the Alfv\'en speed is much lower \citep{2013takasao}.

Many observations show clear signatures of plasmoids formed during the reconnection process, and track their motions using emission signatures. Plasmoids have been observed as periodic blobs in optically thin AIA lines \citep{obsPlasm}.
These plasmoids can appear in the nonlinear evolution of current sheets that are liable to linear resistive tearing modes. Since plasmoids form on the current sheets that also develop Petschek-type configurations with outflows{, once they are formed}, it is relevant to study the linear stability of a CS in the presence of outflows. In an early 2D purely linear MHD analytical study, it has been shown that outflows have a stabilizing effect for the tearing mode \citep{stabOutflow}. In this paper, we will discuss stability aspects of a CS due to tearing in a two-fluid setting. Our simulations contain plasmoids, and we will make synthetic images that directly relate to the observed blob features.

Since our work is using a plasma-neutral two-fluid model, several works that looked into two-fluid reconnection are of direct relevance to our study.
2D simulations in the two-fluid approach show that the reconnection rate is increased because of recombination and larger outflows \citep{slavaRec1}. 
{ Ionization/recombination processes would put additional constraints on the background stratification, leading to nontrivial equilibrium conditions \citep{2021ben}.
A non-static equilibrium introduces a new free parameter through the gradient in the vertical velocity, moreover  it can explain the formation and the properties of the transition region \citep{2023song}.}
In this paper, we generalize the work to stratified settings, but will not include ionization/recombination processes in our simulations.
\cite{slavaRec2} obtain high reconnection rates around 0.1 due to two-fluid effects, which otherwise would be obtained by using Hall, kinetic effects or localized resistivity.
{In stratified setups that were liable to the Rayleigh-Taylor instability, secondary reconnection events} showed that two-fluid effects are locally very important \citep{recBeatrice}. \cite{murtas} found that
plasmoid coalescence happens faster in the two-fluid model than in MHD, because the effective Alfv\'en speed
based on a two-fluid density defined by the collisional coupling, is larger. In 1D two-fluid simulations of slow magneto-acoustic shocks (of the type relevant in a Petschek-type reconnection region), frictional heating leads to a localized region around the `reconnection' point with increased temperature in the neutral fluid \cite{andrewShocks1}. As a consequence a blast wave in the neutral fluid develops
with overshoots in the neutral density and velocity \citep{andrewShocks1,andrewShocks2}. We will show in our multi-dimensional setup how the detailed CS structure may show intricate decoupling (and runaway) effects in the Petschek-type configuration.
Here, we extend previous works by studying reconnection in a 2D stratified setup, accounting for the presence of coupled plasma-neutral species.

We present the numerical setup in Section~\ref{sec:setup}, 
the results of our simulations in Section~\ref{sec:results}.
We then create synthetic views from the simulation snapshots, presented in Section~\ref{sec:synv} and we summarize our conclusions in Section~\ref{sec:summary}. 

\section{Numerical setup}
\label{sec:setup}
\begin{figure}
\FIG{
\includegraphics[width=0.5\textwidth]{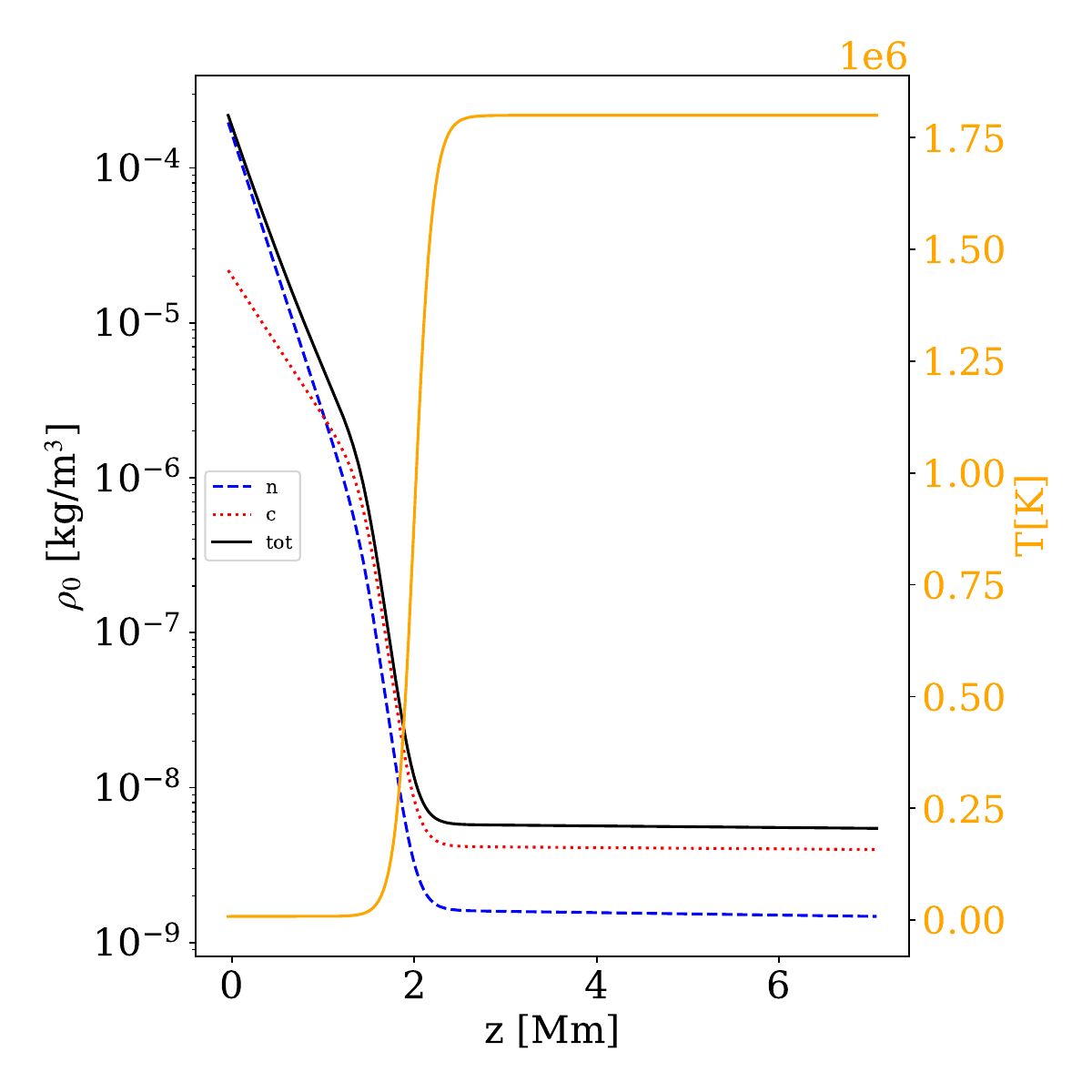}
}
\caption{
Initial profiles as a function of height.
Densities: neutrals (blue dashed line), charges (red dotted line) and total (black solid line) on the left axis.
Temperature (solid orange line) on the right axis.
}
\label{figsetup}
\end{figure}
\begin{figure}
\FIG{
\includegraphics[width=0.5\textwidth]{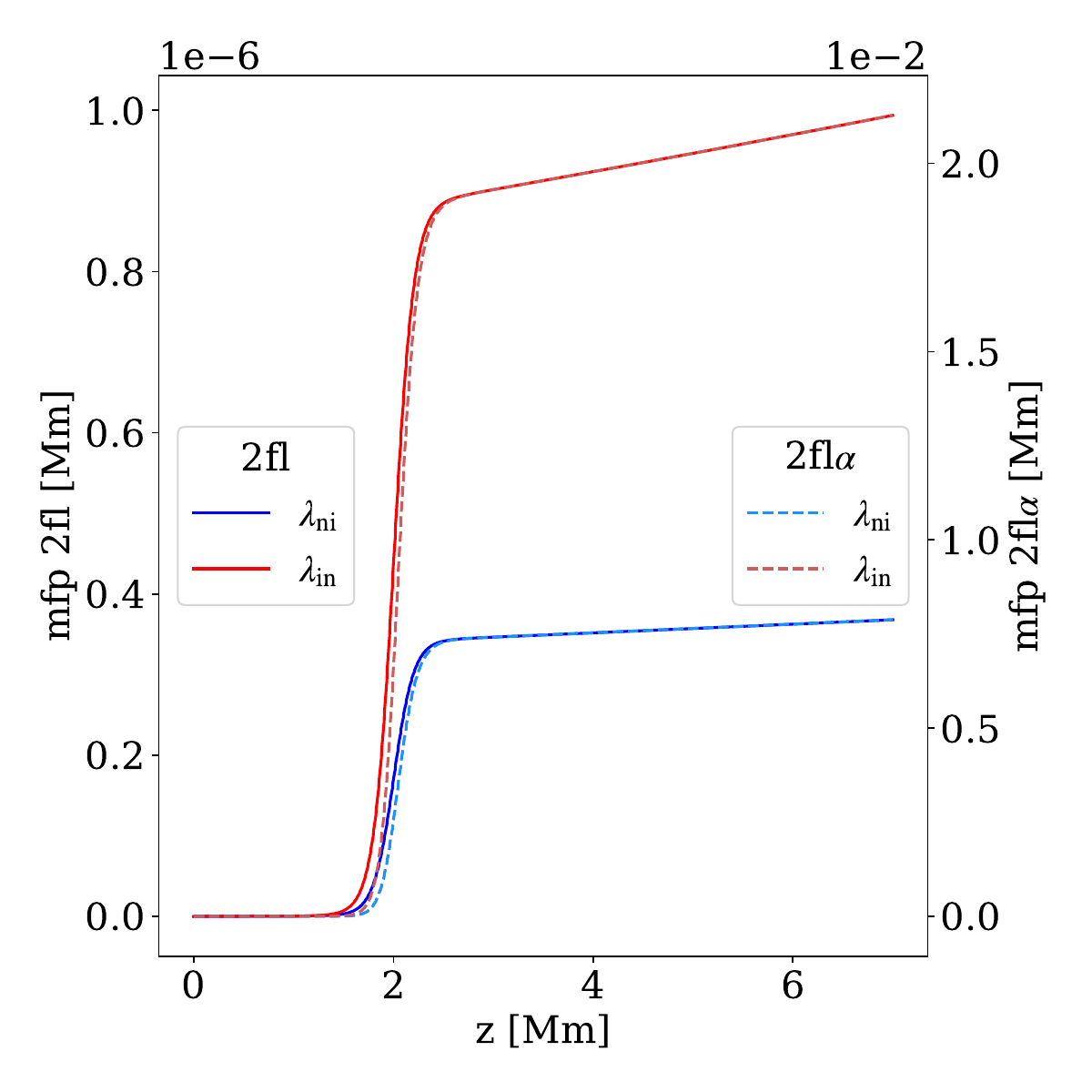}
}
\caption{
Collisional mean free paths as functions of height. We show the mean free paths between ions and neutrals ($\lambda_{\rm in}$, red lines) and between neutrals and ions ($\lambda_{\rm ni}$, blue lines)
for the two collisionality regimes considered: self-consistent from plasma-neutral parameters (2fl, solid lines, left axis)  and reduced case 2fl$\alpha$ (dashed and dotted lines, {right} axis, {note the different order of magnitude}).
The collisional mean free paths are calculated according to Eq. (\ref{eq:mfp}).
}
\label{figsetup2}
\end{figure}

We consider a gravitationally stratified atmosphere where we define a temperature profile  with height $z$ described by:
\begin{equation}
T(z)=T_{\rm ch} + \frac{T_{\rm co}-T_{\rm ch}}{2}\left[\text{tanh}\left( \frac{z-z_{\rm tr}}{w_{\rm tr}} \right) +1\right]\,,
\label{eq:temp_profile}
\end{equation}
where $w_{\rm tr} = 0.2\text{ Mm}$, the transition region height is
    $z_{\rm tr} = 2\text{  Mm}$, the chromospheric and coronal temperature is set through
    $T_{\rm ch} = 8\times10^3\text{ K}$ and $
    T_{\rm co} = 1.8\times10^6\text{ K}$.
The initial profiles of temperature and the neutral and charged, as well as total fluid densities are shown in Figure~\ref{figsetup}. 
We use an ideal equation of state for both neutrals and charges and the normalized mean molecular weight is considered uniform and constant, having the values of 0.5 for charges and 1 for neutral species \citep{2flpaper}.

We use a force-free sheared magnetic field,  changing its far-field (large horizontal coordinate $|x|$) vertical orientation near $x=0$, with uniform magnitude $B_0=10^{-3}~T$, given by
\begin{equation}
  B_{\rm z0} = - B_0 \text{ tanh}\left( \frac{x}{L_s}\right)\,,\quad
  B_{\rm y0} = B_0 \text{ cosh}^{-1}\left( \frac{x}{L_s}\right)\,,
\end{equation}
where $L_s=0.02$~Mm. A similar force-free profile has been used 
in the simulations of \cite{wenzhi2,wenzhi}.
The current sheet width calculated as the full width at half maximum (FWHM) of the corresponding current density component $J_{\rm y0}$ is 0.036 Mm.

We will use a localized resistivity of the particular form
\begin{equation}
\eta(x,z)=\left(\eta_0-\eta_1\right) \text{ exp}\left( -\frac{x^2}{2 L_s} - \frac{(z-z_{\rm rec})^2}{2 L_s}\right) + \eta_1\,,
\end{equation}
where $\eta_0\approx8$~$\Omega m$ and $\eta_1\approx 0.8$~$\Omega m$.
The reconnection point will be at $x=0$ always, but we will vary the reconnection height $z_{\rm rec}$, as mentioned 
in Section~\ref{setup:params} below.
To trigger the reconnection, we adopt an initial velocity variation. The initial perturbation is in the $x$-direction only, trying to bring the field lines closer around the reconnection point, having the form: 
\begin{eqnarray}
  v_x(x,z; t=0) = -V(x,z), \text{for} \,\,x>0\,,\nonumber\\
  v_x(x,z; t=0) = +V(x,z), \text{for} \,\,x<0\,,
\end{eqnarray}
with
\begin{equation}
V(x,z) = A v_A(z) \text{ exp}\left( -\frac{x^2}{2 L_s} - \frac{(z-z_{\rm rec})^2}{2 L_s}\right)\,,
\end{equation}
and
\begin{equation}
\label{eq:va}
v_A(z; t=0)=B_0/\sqrt{\rho_{\rm tot}(z)}\,,
\end{equation}
the total Alfv\'en speed, { having $\rho_{\rm tot}=\rho_{\rm n}+\rho_{\rm c}$ the total density}. We choose the amplitude $A=10^{-1}$.
In a two-fluid simulation, this initial perturbation is the  same for the velocity of charges and neutrals.

The two-fluid model uses the newly developed module in the fully open-source {\tt MPI-AMRVAC} code \citep{2flpaper,mpiamrvac3}.
{ The equations solved are the nonlinear, compressible, resistive two-fluid Eqs.~(1)-(7) from \citep{2flpaper}.}
In a 2.5D geometry the domain $xz$ with $x\in[-0.5,0.5]$~Mm and $z\in[0,7]$~Mm is covered by a grid with base resolution of 256$\times$1024 
and we use five levels of refinement, having an effective resolution of 1024$\times$4096 points and the size of the finest cell $\Delta x$=$9.76\times10^{-4}$~Mm
and $\Delta z$=$1.7\times10^{-3}$~Mm. 
The bottom boundary in the $z$-direction is closed (antisymmetric for vertical velocities and symmetric for the rest of the variables) 
 and we use open boundary conditions (symmetric for all the variables) for the top boundary and both side boundaries in the $x$-direction.
The region $-0.2\le x\le 0.2$ is always refined at the highest level, so that the CS is properly resolved. 
The refinement criterion is based on density only for the MHD cases and equally on charged and neutral density in the two-fluid runs. We use the splitting of the equilibrium force-free magnetic field \citep{B0split} and the gravity stratification for both neutrals and charges \citep{nitin,2flpaper}.
{
In this approach, the magnetic field $\mathbf{B}$, densities $\rho_{\rm n}$, $\rho_{\rm c}$ and pressures  $p_{\rm n}$ and $p_{\rm c}$ are split into a time-independent, 
$\mathbf{B_0}$, $\rho_{\rm n0}$, $\rho_{\rm c0}$, $p_{\rm n0}$, $p_{\rm c0}$ and time-dependent $\mathbf{B_1}$, $\rho_{\rm n1}$, $\rho_{\rm c1}$, $p_{\rm n1}$, $p_{\rm c1}$ parts.
The equations solved in the code are for time-dependent quantities, while the equilibrium conditions are explicitly removed from the equations:
\begin{equation}
\mathbf{J_0}\times\mathbf{B_0}=\mathbf{0}\,, -\mathbf{\nabla}{p_{\rm c0}}-\rho_{\rm c0} \mathbf{g} =0\,,-\mathbf{\nabla}{p_{\rm n0}}-\rho_{\rm n0} \mathbf{g} =0\,.
\end{equation}
Mathematically, the split equations are equivalent to the unsplit equations, but numerically, the splitting helps avoiding an unwanted evolution due to numerical dissipation of the equilibrium. 
}

\subsection{Coupling aspects}
Because of the very low mass of electrons compared to ions, the collisions between charges and neutrals are effectively collisions between ions and neutrals,
and the mean free path between ions and neutrals and between neutrals and ions can be defined as:
\begin{equation}
\label{eq:mfp}
\lambda_{\rm in} = \frac{v_A}{\nu_{\rm in}}\,;\quad \lambda_{\rm ni} = \frac{v_A}{\nu_{\rm ni}}\,,
\end{equation}
where $\nu_{\rm in}=\alpha\rho_n$ and $\nu_{\rm ni}=\alpha\rho_c$ denote collision frequencies between ions and neutrals and between neutrals and ions, respectively \citep[see Eq.~(15) in ][]{2flpaper}.
The characteristic velocity is the Alfv\'en speed of the whole fluid, as defined (generalized to $v_A(x,z,;t)$) by Eq.~(\ref{eq:va}) 
\citep[see also Eq.~(37) in ][where the characteristic velocity was chosen as the fast magneto-acoustic speed]{2flpaper}.
The collisional parameter $\alpha$ \citep[see Eq.~(A.3) in][]{2flpaper} { is defined as:
\begin{equation}
\alpha = \frac{2}{{m_H}^{3/2} \sqrt{\pi}} \sqrt{ k_B T_{\rm cn}}  \Sigma_{in} 
\end{equation}
where the collisional cross-section considered here is $\Sigma_{in} = 10^{-19} \text{m}^{2}$.
$T_{\rm cn}$ is the average of temperatures of the neutrals and charges. }
In this paper, we consider two cases for the collisional coupling, one when $\alpha(x,z;t)$ is calculated consistently from instantaneous plasma parameters and another
where it is set to a constant value. 
When $\alpha$ is calculated self-consistently, its profile varies slowly with height, with values initially between $5.8 \times 10^{12}$~m$^3$/kg/s and $8.2 \times 10^{12}$~m$^3$/kg/s. These minimum and maximum values remain practically unchanged at the end of the simulation. These high values imply that the coupling is near perfect (and hence MHD-like behavior is expected) throughout the domain. We will compare this to a run where we instead set $\alpha$ to a constant value throughout. This constant value of $\alpha=3.84 \times 10^8$~m$^3$/kg/s is almost four orders of magnitude smaller than the consistently computed values. However, as we argue in the section below, this reduced coupling value ends up to be more representative for actual solar settings. Moreover, in that regime, we will have two-fluid effects that are important in stratified reconnection setups.

    \subsection{Parameters of the simulations}\label{setup:params}

For our study of reconnection in stratified settings, we will compare three collisional regimes, namely: (1) a single fluid
     MHD model (label ``MHD''); (2) a
    two-fluid plasma-neutral model where the collisional parameter $\alpha$ is calculated self-consistently from plasma values (label ``2fl''); (3) a 
    two-fluid model where the collisional parameter $\alpha$ is constant and set to a smaller value of $3.84 \times 10^8$~m$^3$/kg/s (label ``2fl$\alpha$'').   We will study 6 cases in total, since each collisional regime will vary the reconnection point from $z_{\rm rec}=2$ Mm, with reconnection in the upper chromosphere to low transition region, to $z_{\rm rec}=4.5$ Mm, for coronal reconnection.

{    
In the MHD model, the initial equilibrium atmosphere is constructed by summing the densities (shown in Figure \ref{figsetup}) and pressures for charged and neutral fluid at all heights.
}
  The resulting mean free paths between ions and neutrals and between neutrals and ions for the two two-fluid models, 2fl and 2fl$\alpha$,
 are shown in Figure~\ref{figsetup2}. In this 2fl$\alpha$ case, both values of the mean free path in the transition region and above are ${\cal{O}}(0.01 \text{Mm})$, hence they are similar 
to the width of the CS, while in the 2fl case the values of the mean free path are ${\cal{O}}(10^{-6} \text{Mm})$. Because of the weak dependence on height of  the value of $\alpha$ calculated self-consistently, both profiles of the mean free paths (2fl and 2fl$\alpha$) look similar, being dominated by the variation of the density, included in the calculation of the mean free path as from Eq.~\ref{eq:mfp}.
{ The mean free path does not change significantly during the simulation, being rather determined by the equilibrium plasma parameters.}

We will now argue that the 2fl$\alpha$ case is more solar-relevant. Indeed, we used a temperature profile defined by Eq. \ref{eq:temp_profile}, which is slightly smoother than VALC \citep{VALC}, but it has the advantage that we directly control the width ($w_{\rm tr}$) of the transition region (TR). A similar temperature profile defined by an analytic function 
 has been used by other authors \citep{tempProfile}. This overly smooth temperature variation also implies a larger fraction of neutrals 
at both used reconnection heights, especially at the coronal reconnection point
$z_{\rm rec}=4.5$~Mm, taking into account that the scale height of the neutrals is twice smaller than that of the charged particles.
The other reconnection case has $z_{\rm rec}=2$~Mm, i.e. starts its reconnection at the middle of the TR. How effective the collisional coupling between plasma-neutral species is at these heights, is largely set by the densities they attain there.

When we integrated the vertical profiles, at the base of the atmosphere at $z=0$,
we used the total number density from the VALC model namely $n_T\approx10^{23}$~m$^{-3}$, but we had to consider an ionization fraction of 0.1,
so that we still obtain more charges in the corona despite the smoothing of the temperature profile.
Hence, the adopted temperature variation, together with the imposed bottom densities lead to a reversal of the dominance of neutrals over charges at $z=1$~Mm, while the entire transition region and corona is charge-dominated.
We find that the overly smoothed temperature profile actually increased the 
 density { in the upper chromosphere and corona} above observed  values, making collisional coupling now larger there. This justifies the use of a smaller and more representative value of $\alpha$ in the simulation 2fl$\alpha$, which would mimic the actual coupling found for the normally smaller densities. Indeed, the mean free path between ions and neutrals in the two-fluid reconnection simulations of \cite{slavaRec2}, which was self-consistently calculated, was about 100~m. This is nicely situated between our 2fl and 2fl$\alpha$ case.
Moreover, there are different recipes for calculating collisional frequencies, due to different values for the collisional cross sections available in the literature, which lead to large differences \citep{2017sykora}.
The very small mean free path in our 2fl case suggests that it can be considered an MHD limit case, and this will be confirmed in our simulations below.

\section{Results}
\label{sec:results}

\begin{figure*}
\FIG{
\includegraphics[width=0.5\textwidth]{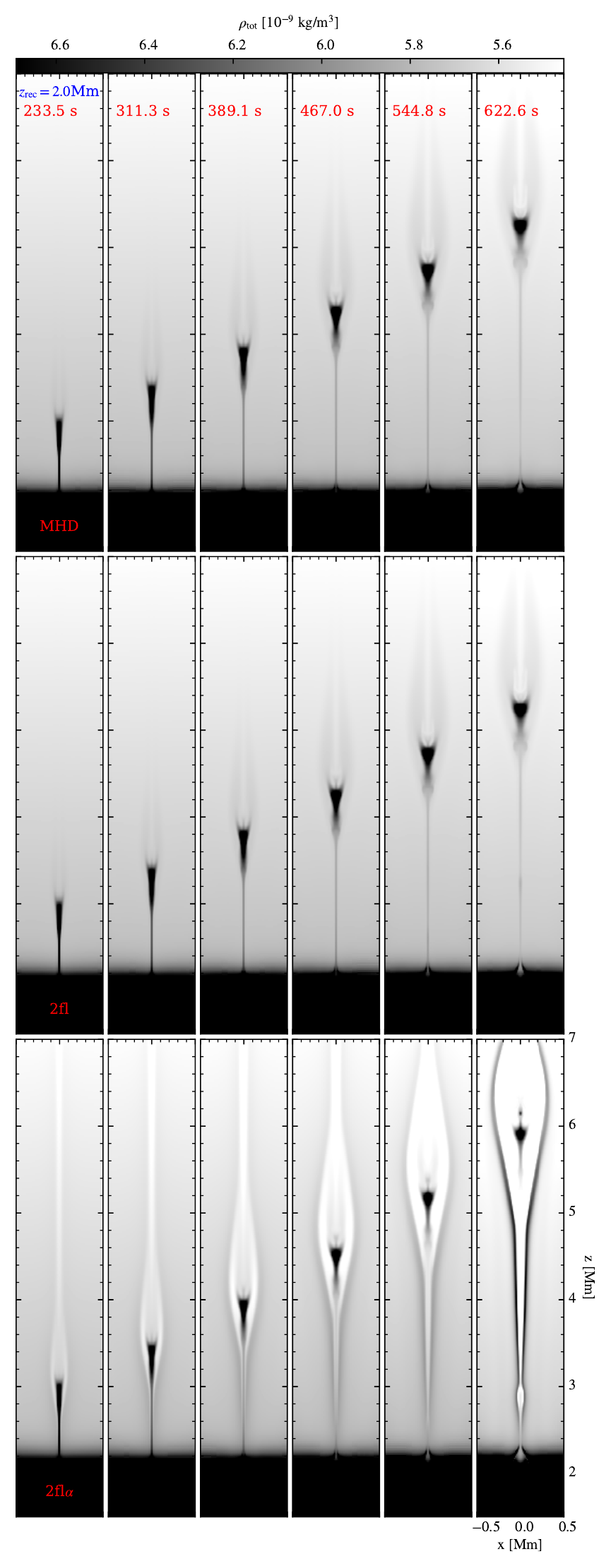}
\includegraphics[width=0.5\textwidth]{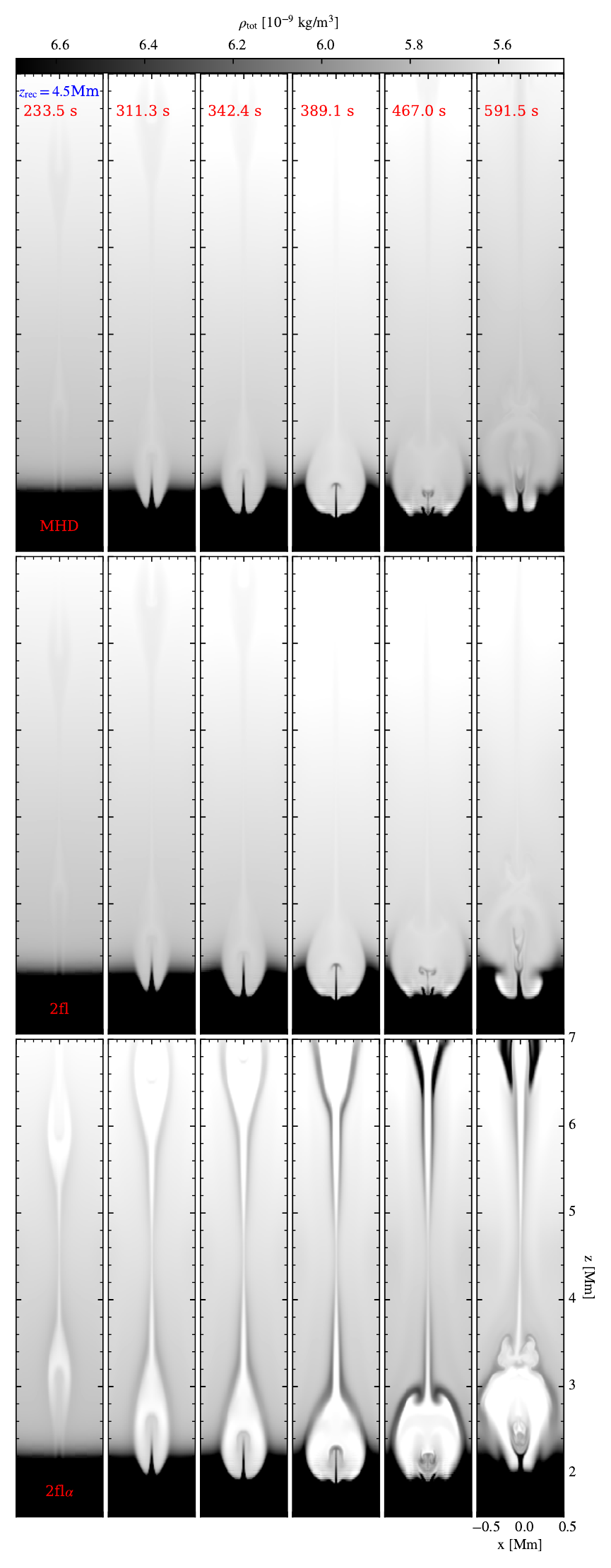}
}
\caption{
Time evolution of total density for all six simulations. Left panels: $z_{\rm rec}=2$ Mm.
Right panels: $z_{\rm rec}=4.5$ Mm. A different physics model is shown in each row: top row: MHD; middle row: 2fl; bottom row: 2fl$\alpha$.
An online animation (case z2.0-2flalpha.mp4 or 2fl$\alpha$) overplots the adaptive grid for the neutral density.
}
\label{fig:overview}
\end{figure*}
\begin{figure}
\FIG{
\includegraphics[width=\columnwidth]{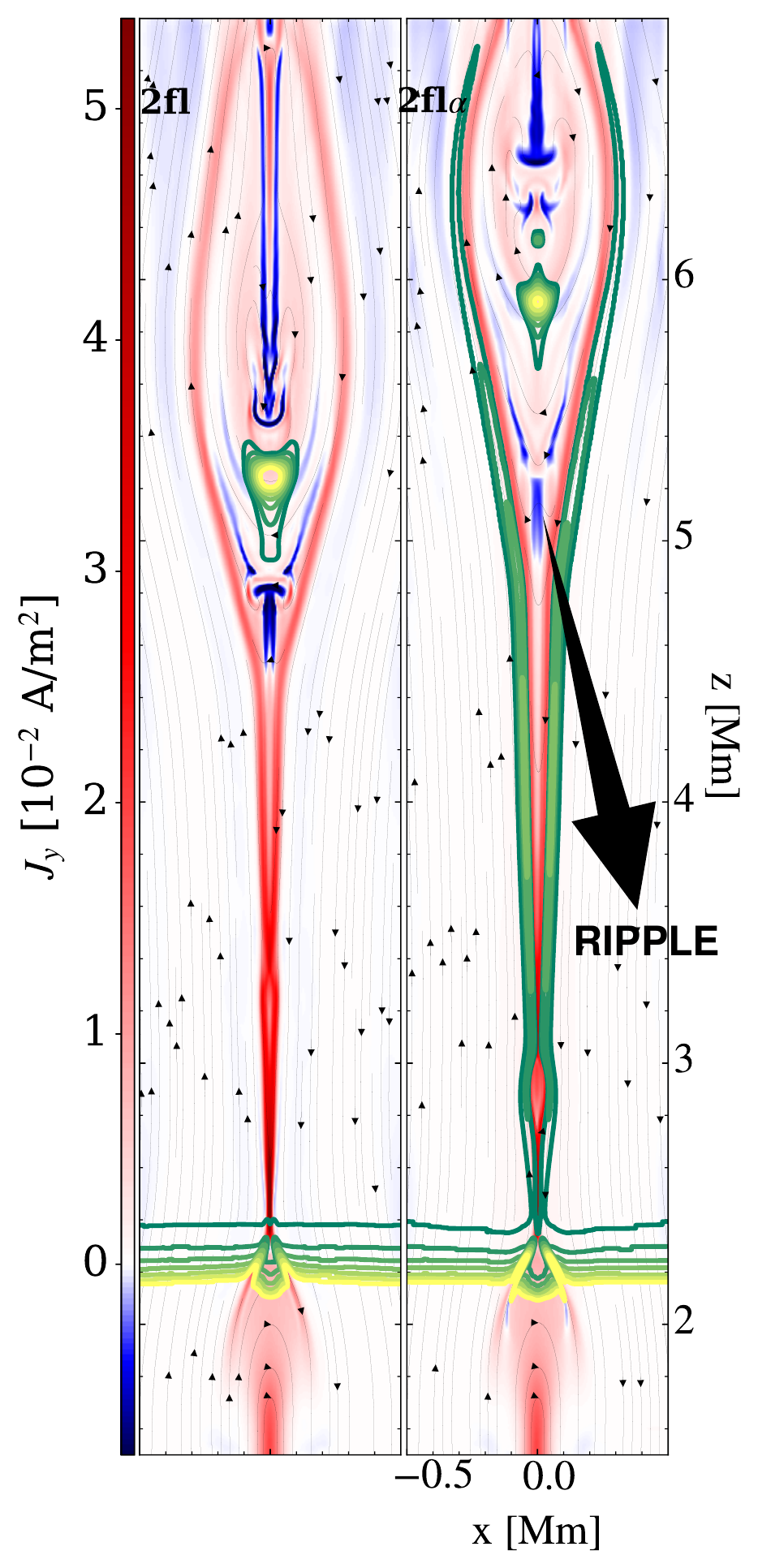}
}
\caption{
Comparison between the out of plane current density. For two-fluid models: 2fl (left) and 2fl$\alpha$ (right) at $t=622.6$ s. Black lines with arrows are magnetic field lines, yellow-to-green colored contours relate to the density variation.
{
The ripple in the magnetic field, associated with the reversal in $B_z$ sign, located at z$\approx$5~Mm is indicated in the figure.
}
}
\label{fig:compjy}
\end{figure}

\begin{figure}
\FIG{
\includegraphics[width=\columnwidth,height=9cm]{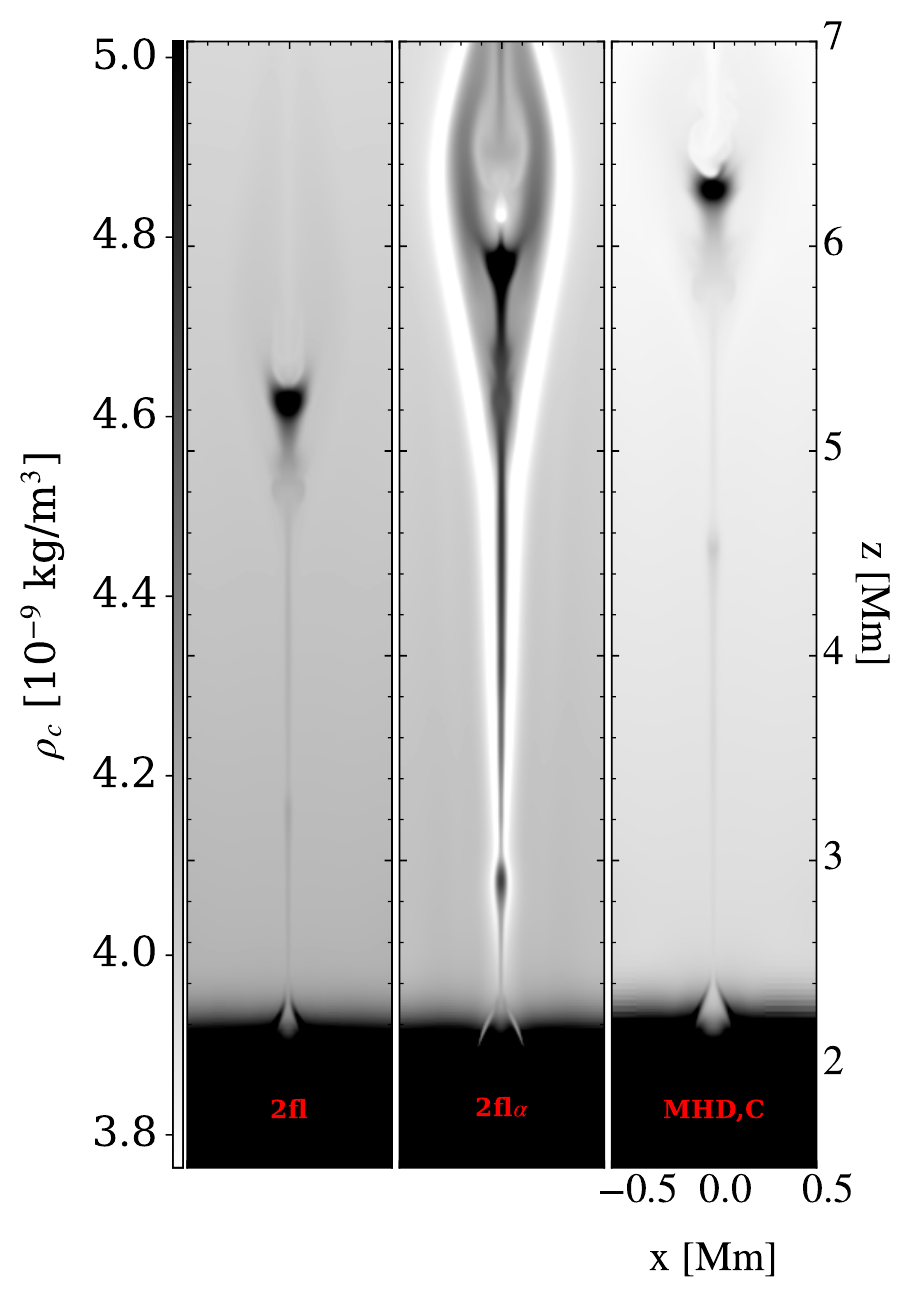}
\includegraphics[width=\columnwidth,height=10cm]{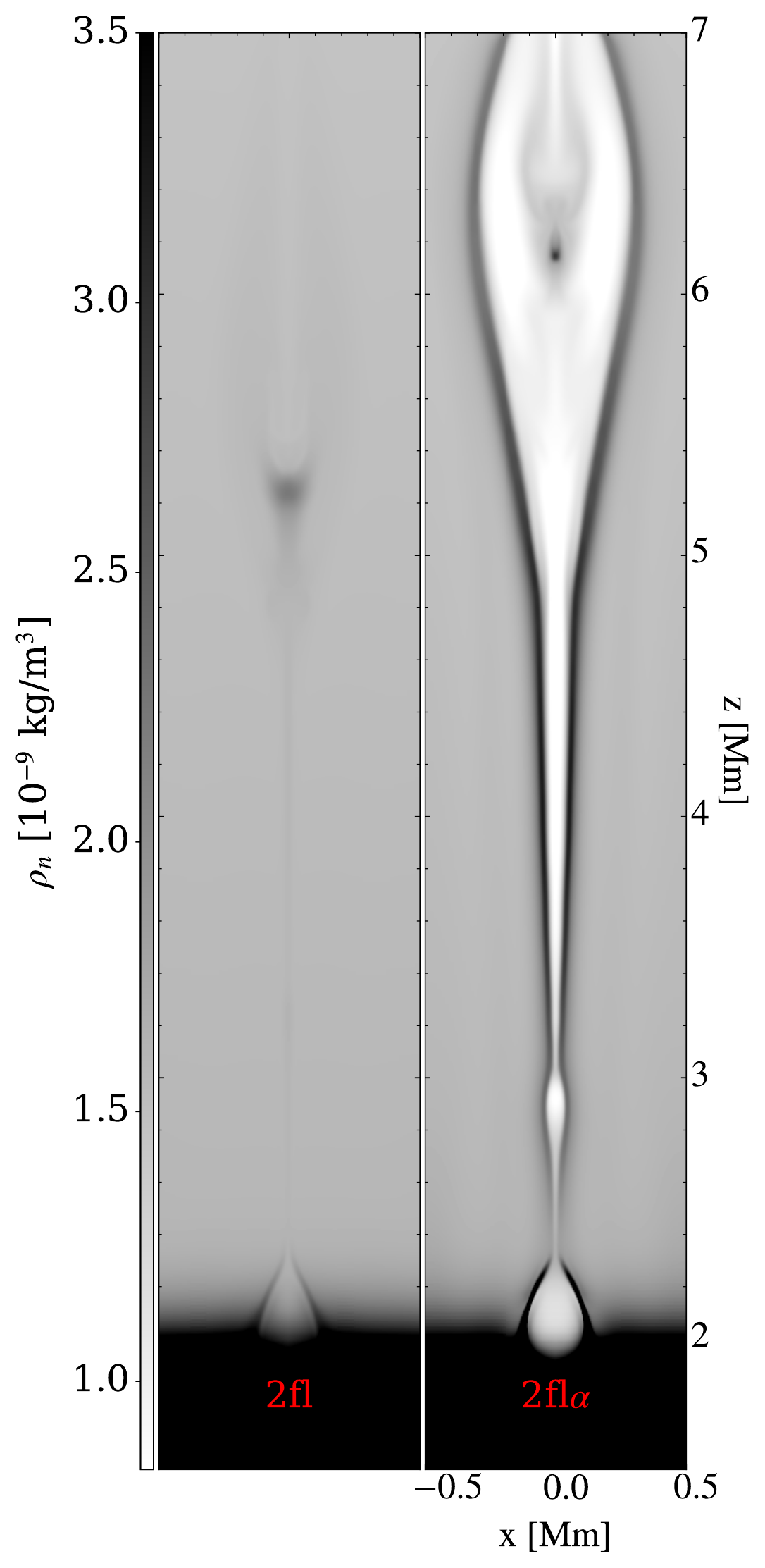}
}
\caption{
Comparison between the charged densities (top panels) and the neutral densities (bottom panels) for $t=622.6$ s.
Top row: 2fl (left panel), 2fl$\alpha$ (middle panel), MHD with charged fluid only (right panel). 
Bottom row: 2fl (left panel), 2fl$\alpha$ (right panel). 
}
\label{fig:comprho}
\end{figure}
\begin{figure}
\FIG{
\includegraphics[width=\columnwidth,height=9cm]{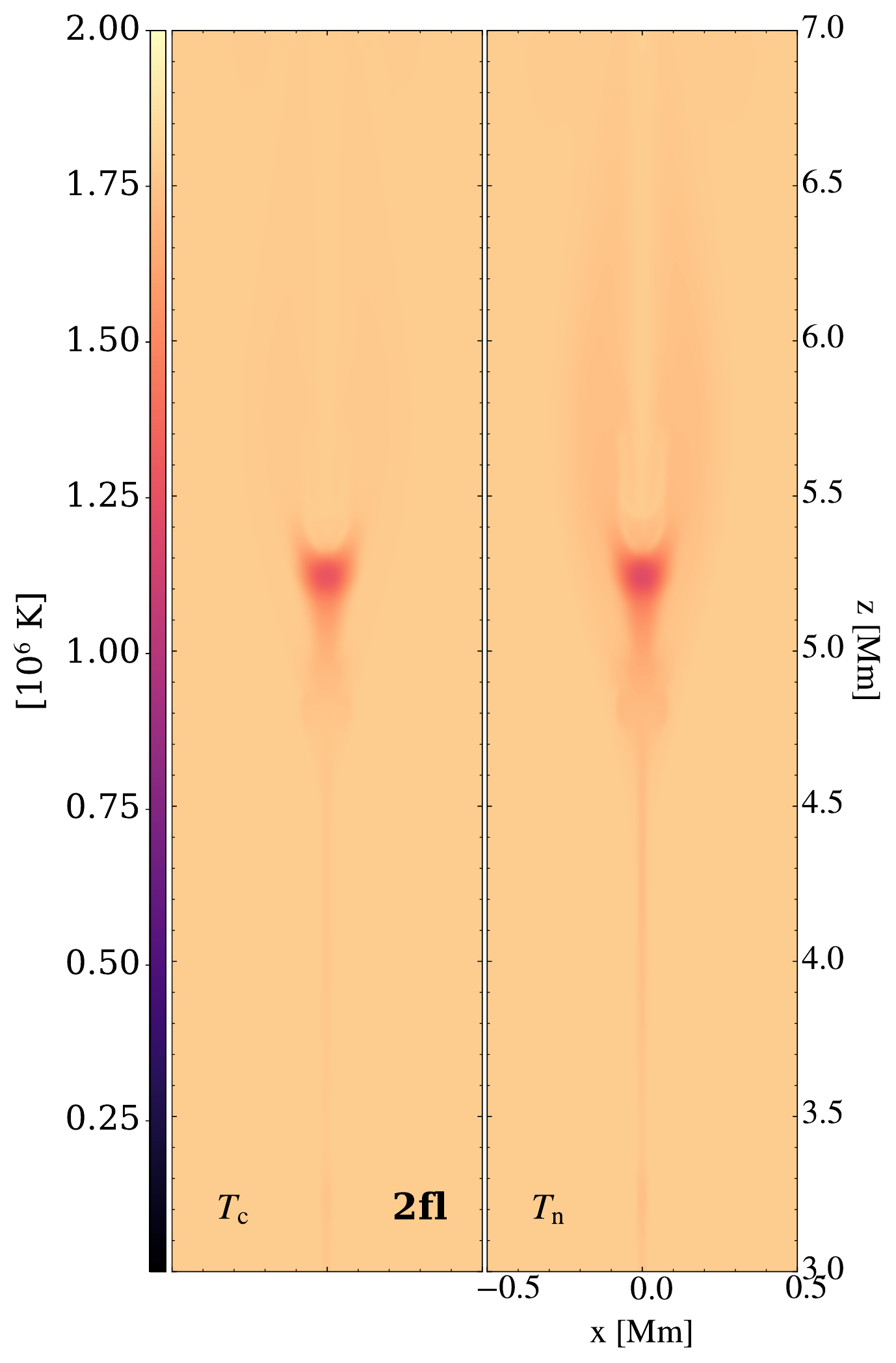}
\includegraphics[width=\columnwidth,height=9cm]{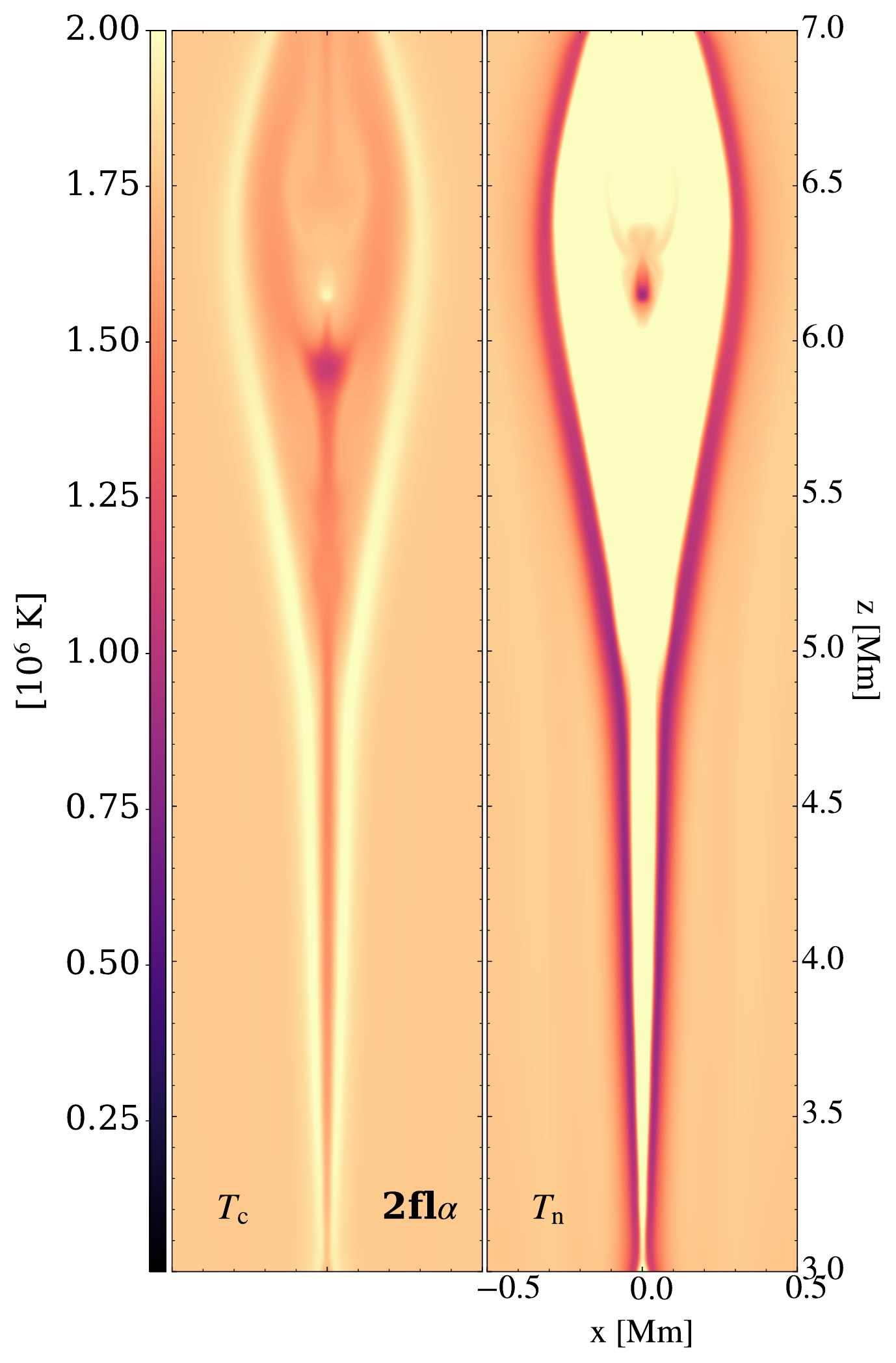}
}
\caption{
Comparison between the temperatures of neutrals and charges at $t=622.6$ s.
Top panels: 2fl, bottom panels: 2fl$\alpha$.
}
\label{fig:comptemp}
\end{figure}

In all six cases considered (MHD, 2fl, 2fl$\alpha$ for the coronal versus TR reconnection case) the reconnection develops, producing bidirectional outflow jets, traveling away from the reconnection point, which can be seen in the snapshots of the total density 
shown in Figure~\ref{fig:overview}.
As the atmosphere is gravitationally stratified, the jets traveling upwards are denser and those traveling downwards are less dense than the surrounding
fluid located at the same height. 
The figure shows that for TR reconnection, we get a pronounced upwards jet that is accompanied by a vertically oriented CS that lengthens as time progresses.
From the time evolution of the upwards moving jets in the $z_{\rm rec}=2$~Mm case we can estimate a vertical velocity of $\approx$ 8~km/s. An online animation for this case (z2.0-2flaplha.mp4) overplots the adaptive grid for the neutral density.  In the case of coronal reconnection, we find a clear two-sided (up-down) jet forming, where the lower one ultimately interacts with the TR and chromosphere (forming post-flare loops). 
For both reconnection heights considered, $z_{\rm rec}=2$~Mm and $z_{\rm rec}=4.5$~Mm, the MHD (top row) and the 2fl models (middle row) give very similar results, as expected according to the collisionality regime.
A clear difference appears in the snapshots for the 2fl$\alpha$ model (bottom row of Figure~\ref{fig:overview}), with the development of a less dense region with an edge of increased density, and this is seen in both $z_{\rm rec}$ cases. 
In order to better understand these differences, we further analyze 
snapshots for $z_{\rm rec}=2$ Mm at $t=622.6$ s, the last time shown in Figure~\ref{fig:overview} (left column).

\subsection{Analysis for the case with TR reconnection}

We plot in Figure~\ref{fig:compjy} the out-of-plane current density $J_y$ for the 2fl and 2fl$\alpha$ models, and overplot magnetic field lines and isocontours of total density.
Except for the fact that the 2fl$\alpha$ seems a bit further evolved in time (as the top magnetic island is located a bit higher), the current density structures are very similar for the two models. 
We observe that the dense edges in the 2fl$\alpha$ case are located just outside the current sheets.
{ Because of the localized resistivity, the simulations evolve towards a Petschek-like reconnection.}
Slow shocks that accompany the reconnection,
 traveling in the $x$-direction, widen gradually the distance over which $B_z\approx 0$, splitting the CS into two current sheets, seen as ``V'' structures in the images.
Theoretically, in an assumed MHD stationary Petschek reconnection state (without stratification), the slow shock discontinuity is located along this V-pattern,  
$B_z$ is exactly zero inside the V and $B_x$ is uniform within. Hence, the two current sheets should actually be infinitely thin in the ideal MHD limit that holds away from the resistive reconnection layer.
However, in our more realistic setup we find that it is possible for $B_z$ to locally reverse sign, creating another current sheet with $J_y$ of opposite sign, as seen most clearly in the snapshot
for the  2fl$\alpha$ case at height $z\approx5$~Mm, at the location of the corresponding ripple in the magnetic field line.
This reversal in the sign of $B_z$ {was in 1D two-fluid models associated with a reversal in the sign of $v_x$ and was found related to} the heating produced by the two-fluid effects \citep{andrewShocks1,andrewShocks2}.
However, this reversal in the sign of $B_z$ is observed in our MHD (and 2fl) simulations as well, so that the Ohmic heating might be the cause in our simulations.  Note that the magnetic field lines in Figure~\ref{fig:compjy} show the expected post-flare loop configuration below the reconnection site. The 2fl$\alpha$ case also shows a plasmoid structure forming { at later stage}, 
and we will discuss this in Section~\ref{sec:plasmoid}.

We compare snapshots of density of charges and neutrals, separately for the two models in Figure \ref{fig:comprho} (note that our earlier Figure~\ref{fig:overview} showed total densities). 
We observe that in the 2fl case the density of charges and neutrals have similar structures, the difference being mainly in magnitude.
{ The magnitude scales with the background density, the neutral density being smaller in the corona than the charged density.
The neutral density shows  more variation with height, since the scale height of the neutrals is half of that of the charges.}
In contrast, in the 2fl$\alpha$ case, we observe a clear reversal of contrast, namely a central region with low neutral density and high charged density,
surrounded by a shell of high neutral density and low charged density.
{ We also show in the charged density plots the density snapshot of a MHD run where we used the charged fluid properties only, this corresponding to the zero collisions $\alpha=0$ limit. 
The snapshot seems more evolved in time than the previous cases: 2fl and 2fl$\alpha$. A smaller (effective) density implies a higher Alfv\'en speed and shorter hydrodynamic timescales, however the edge of increased neutrals is rather related to an incomplete coupling regime, and not to a smaller effective density. Similarly to the conclusion of \cite{murtas}, a smaller effective density due to the collisions is associated to smaller timescales and faster evolution, as we have seen 
previously in Figure~\ref{fig:compjy}.  } 
The same reversal can  be seen in the temperature maps as well for the 2fl$\alpha$ case, as shown in Figure \ref{fig:comptemp}, where the higher density 
regions have smaller temperatures, for both neutrals and charged fluids.
The 2fl case (top panels of Figure \ref{fig:comptemp}), however, shows similar structures in the temperatures of neutrals and charges.
\begin{figure*}
\FIG{
\includegraphics[width=\textwidth]{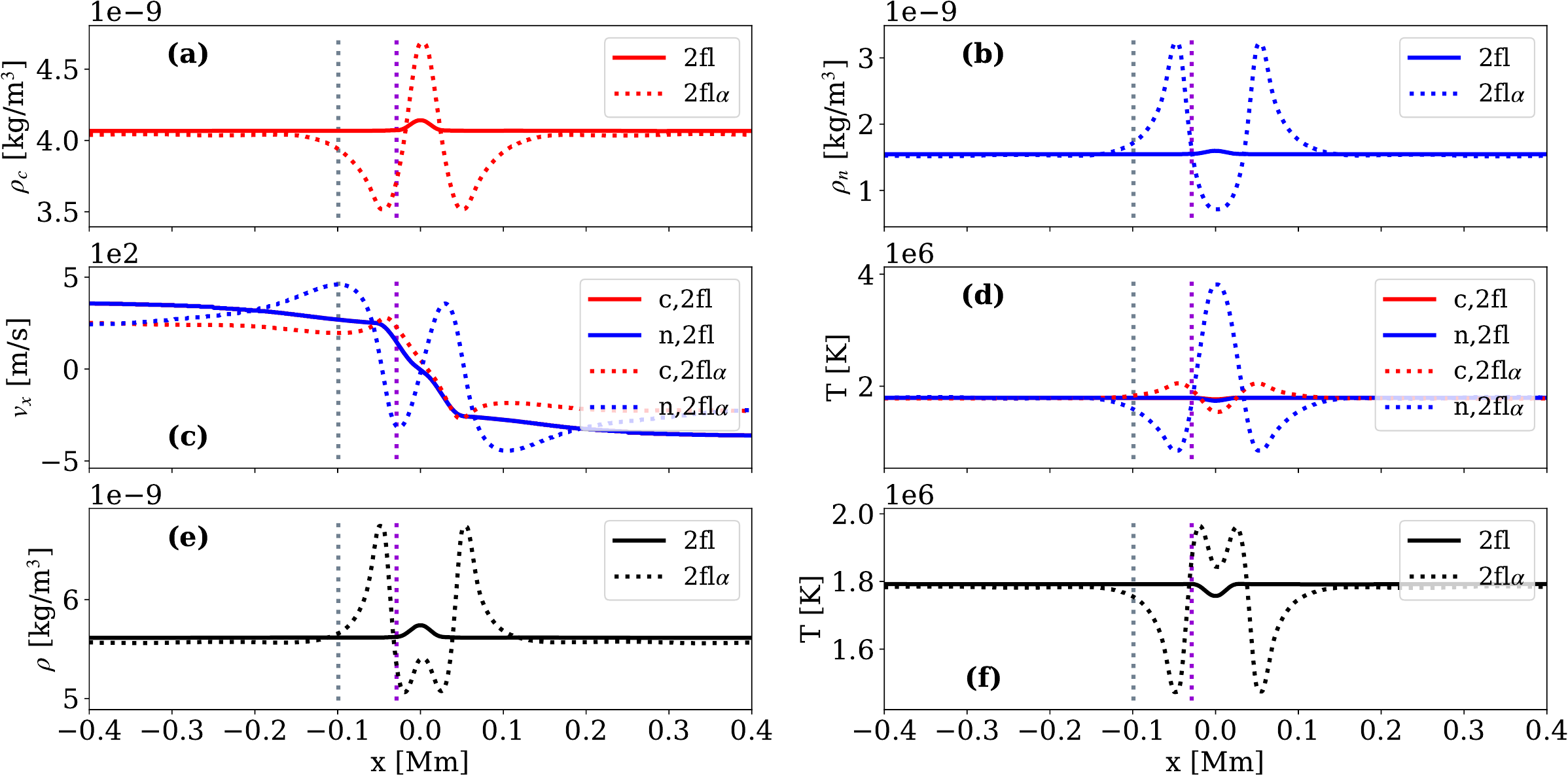}
}
\caption{
Horizontal cut at coronal height $z=4$ Mm at time $t=622.6$ s for the simulations with TR reconnection or $z_{rec}=2$ Mm.
(a) Density of charges, (b) Density of neutrals, (c) $x$-velocity for charges and neutrals, (d) Temperature of charges and neutrals,
(e) Total density, (f) Center of mass temperature from Eq.~(\ref{eq:temp_2fl}).
Different curves are for the two models considered: 2fl and 2fl$\alpha$ and quantities for charges and neutrals are shown in the same plot in panels (c) and (d),
as indicated in the legend.
The primary and secondary maximum peaks in the neutral collisional heating term for case 2fl$\alpha$, as shown in detail in Fig.~\ref{fig-termsq}, located at $z=-0.03$~Mm and $z=-0.1$~Mm are indicated by vertical violet and gray dotted lines, respectively.
}
\label{fig:z2qcut}
\end{figure*}
\begin{figure}
\FIG{
\includegraphics[width=8.5cm]{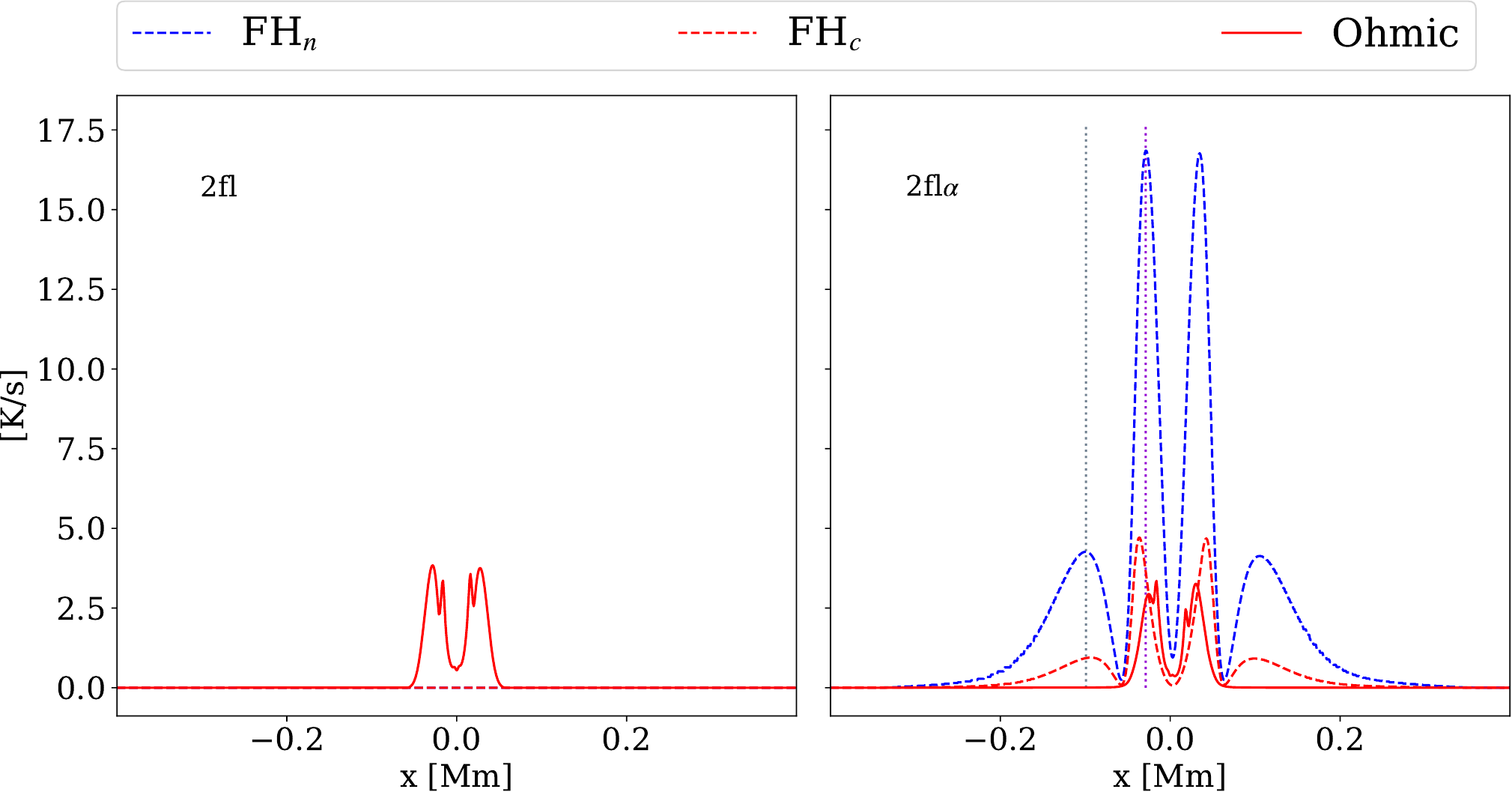}

\includegraphics[width=8.5cm]{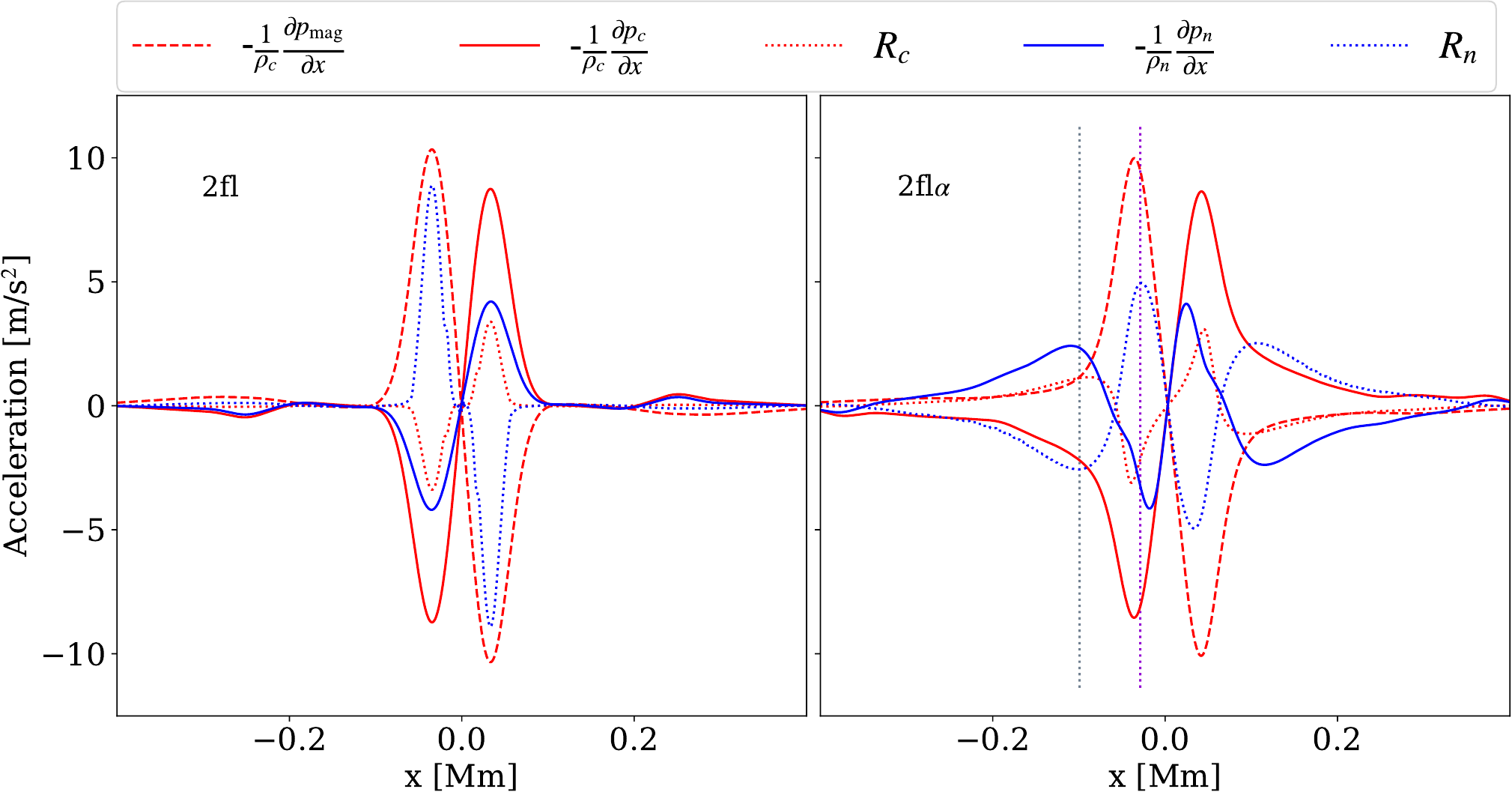}
}
\caption{Horizontal cuts along $z=4$~Mm for the case with TR reconnection or $z_{\rm rec}=2$~Mm at time $t=622.6$ s, the same time as for Figure~\ref{fig:z2qcut}. 
Quantities corresponding to charged fluid are shown with red lines and those
for neutral fluid with blue lines. 
Left panels: 2fl case; Right panels: 2fl$\alpha$ case.
Top row: Terms in temperature equation: Ohmic term (solid red line), collisional heating terms (blue dashed lines: red for charges and blue for neutrals).
Bottom row: Terms in velocity equation;
acceleration due to: magnetic pressure (red dashed line), gas pressure gradient (solid lines: red for charges and blue for neutrals),
collisions (dotted lines: red for charges and blue for neutrals).
The primary and secondary maximum peaks in the collisional heating associated with neutrals, located (and shown only for negative $x$, because of symmetry) at $z=-0.03$~Mm and $z=-0.1$~Mm are indicated by vertical violet and gray dotted lines, respectively for the 2fl$\alpha$ case (right panels). 
}
\label{fig-termsq}
\end{figure}

In order to understand this clear difference in the 2fl$\alpha$ case, which must be caused by two-fluid effects,
 we plot in Figure~\ref{fig:z2qcut} different quantities along a horizontal cut located at $z=4$~Mm.
 These 1D profiles are consistent with the previous 2D images, most notably Figures~\ref{fig:comprho} and \ref{fig:comptemp}, where they cut across the V-shaped current sheet structures discussed earlier. 
For the 2fl model the structures in the charged and neutral density (panels (a) and (b)) are similar. The $x$-velocity (panel (c)) and temperature (panel (d)) profiles overlap for both fluids,
meaning that the two fluids are coupled in both velocity and temperature.
For the 2fl$\alpha$ model, more charges imply less neutrals and higher density implies lower temperature.
The temperature of the neutrals increases by more than $1.6 \times 10^6$ K at the center of the CS.
Further from the center of the CS, the velocities of charges and neutrals are coupled, however inside the CS, they are completely different.
The neutral velocity changes sign at the center of the CS, meaning that the neutrals are going out of the CS. Hence, the 2fl$\alpha$ case which shows a pronounced anticorrelated structure in density and temperature for charges versus neutrals is clearly demonstrating decoupling between both species through the CS structure.

Panels (e) and (f) in Figure~\ref{fig:z2qcut} show the total density, $\rho$, 
and the temperature of the center of mass, defined as
\begin{equation}
T^{\rm 2fl} = \frac{1}{\rho}(\rho_c  T_c + \rho_n T_n)\,.
\label{eq:temp_2fl}
\end{equation}
The total density profile is very similar to the separate neutral and charged density profile for the 2fl model. For this  case,
this positive peak in the density, and a corresponding negative peak in the temperature, are related to the fact that the (vertically flowing)
reconnection outflow comes from a lower height with higher density and lower temperature. However, the two-fluid solution in the 2fl$\alpha$ case, where the collisional effects are enhanced, behaves rather different, and demonstrates a nonlinear runaway effect.
In the case of the 2fl$\alpha$ model, there is a central depletion in total density, where charges accumulate and neutrals deplete, and this is surrounded by a layer of enhanced total density where charges deplete and neutrals accumulate. A tiny central positive peak, of similar magnitude as in the 2fl model, still remains in the very center CS region.  
Seen from an MHD point of view, the whole fluid (both neutrals and charges) heats because of the collisions. This creates a peak in the center of mass temperature (panel (f), dotted line) and the entire fluid expands, so that the total density decreases towards the center of the CS (panel (e), dotted line). In the various panels of Fig.~\ref{fig:z2qcut}, vertical dotted lines show extremal positions in the neutral collisional heating term for case 2fl$\alpha$, discussed in more detail in what follows.

In order to better understand the temperature and velocity profiles seen in Fig.~\ref{fig:z2qcut} we show the terms which enter the temperature (top panel) and velocity (bottom panel) equations in Fig.~\ref{fig-termsq}. The primary and secondary maximum peaks in the neutral frictional heating term (blue dashed line) for the 2fl$\alpha$ case (top-right panel),  are located at a distance of 0.03~Mm and 0.1~Mm away from the center of the CS, and these are indicated by vertical violet and gray dotted lines, respectively (those are repeated in each frame of Fig.~\ref{fig:z2qcut}).
The peak in the Ohmic heating term (red solid line) is located close to the primary maximum peak in the neutral frictional heating term of the 2fl$\alpha$ case (i.e. near the vertical violet dotted line). This frictional heating is negligible in the 2fl case (top-left panel). In the 2fl$\alpha$ case (top-right panel),
the charged fluid frictional heating term (red dashed line) has a similar profile to the neutral frictional heating term, but with smaller values, because of the higher charged density compared to the neutral density. { In the 2fl$\alpha$ case, the neutral frictional heating peak is five times larger than the peak value in the Ohmic heating.}

We now turn attention to what causes the velocity decoupling effects, by discussing the forces as shown in Fig. \ref{fig-termsq}, bottom panels.
The inflow into the CS (as seen in panel (c) of Fig.~\ref{fig:z2qcut})
is driven by the magnetic pressure gradient which acts only on charges, producing initial decoupling in the velocities between neutrals and charges when entering the CS.
This magnetic pressure gradient (red dashed line in bottom panels of Fig.~\ref{fig-termsq}) pushes the charges and the collisionally coupled neutrals inside the CS
against the gradient of their pressures (solid lines). In the 2fl case, only Ohmic heating matters, and the overall decoupling between neutrals and charges stays minimal throughout the CS. 
In the 2fl$\alpha$ case, the neutral and charged fluids decouple in velocity in a more pronounced way throughout the entire CS, again indirectly driven by the magnetic forces which drive the reconnection. But the initial decoupling grows and collisional heating of neutrals causes the neutrals to heat, expand and go out of the CS, accumulating between the primary and secondary heating point (violet and gray lines, respectively), as observed in panel (b) of Fig.~\ref{fig:z2qcut}. 
The inflow of charges there decreases because of the increased collisions with the neutrals that accumulated outside the CS (see the local minimum observed for the dotted red curve at the location of the gray line in panel (c) of Fig.~\ref{fig:z2qcut}).
Therefore, the decoupling in velocity increases further, and also the frictional heating at this secondary neutral heating maximum point (the gray vertical line), which again increases the neutral pressure. 
Thus, the neutral inflow increases  as a local maximum is observed for the dotted blue curve at the location of the gray line in panel (c) of Fig.~\ref{fig:z2qcut}. This is because of the neutral pressure gradient, which shows a local maximum peak in the solid blue curve at the location of the gray line in bottom-right panel of Fig.~\ref{fig-termsq}. This in turn drags the charges into the CS: a local maximum peak is seen for the red dotted line in bottom-right panel of Fig.~\ref{fig-termsq}, which has positive values around the gray line. This implies acceleration of the plasma towards the center of the CS, in the same direction and partially overlapping the magnetic pressure gradient curve. Therefore, the velocity of charges towards the center of the CS is increased and has opposite sign to that of the neutrals, leading to more decoupling and to a runaway nonlinear instability. The heating of the charges at the primary heating point (violet line) slows down in this runaway instability process.
The charges enter faster into the CS because of decreasing collisions with neutrals and this accelerates the process (besides leading to a thinner CS and faster outflows).
The runaway process is further enhanced by the charges expanding due to the (collisional) heating at the secondary point (gray line). 
The runaway process creates a region of accumulation of neutrals bordering the CS (within the hydrodynamic timescale on which the charges enter in the CS due 
to the magnetic pressure gradient), 
which creates a secondary (collisional) heating point whereby the charges get pushed faster into the CS. 
\begin{figure*}
\FIG{
\includegraphics[width=\textwidth, height=9cm]{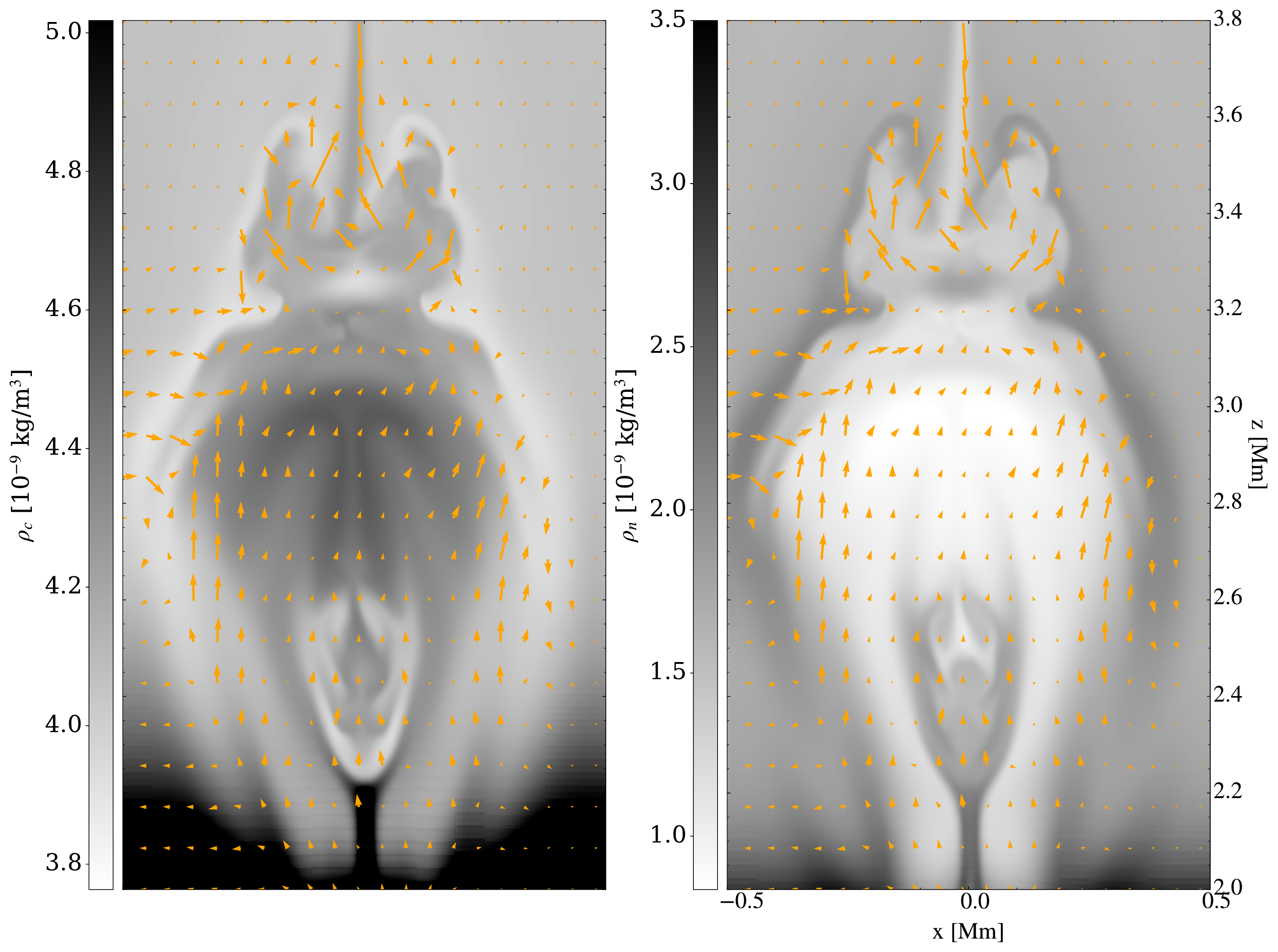}

\includegraphics[width=\textwidth,height=9cm]{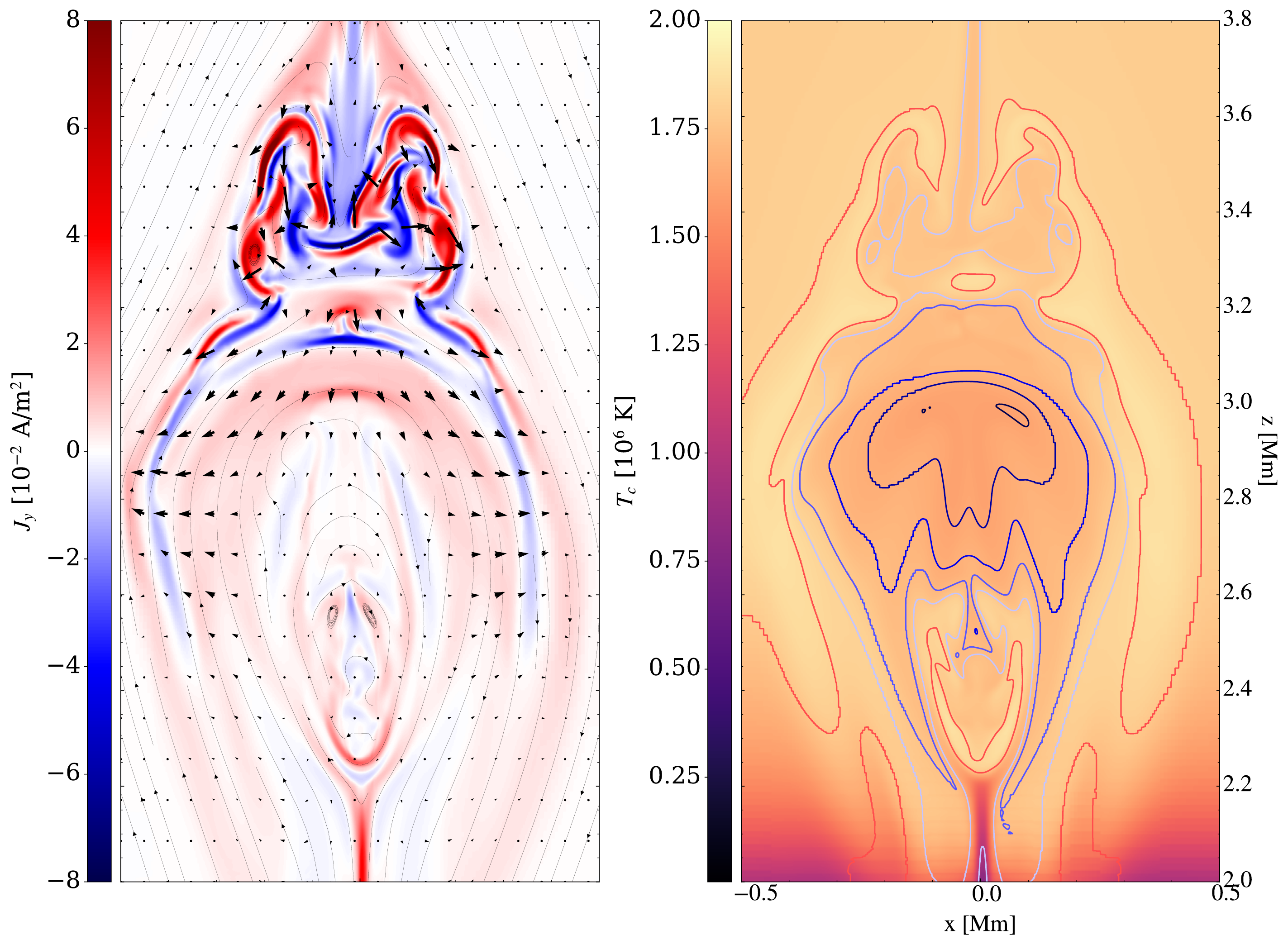}
}
\caption{Two-fluid coronal reconnection downflows impacting the chromosphere.
Snapshots of charged density (top-left), neutral density (top-right), out of plane current density (bottom-left) and charged temperature (bottom right)
for the simulation $z_{\rm rec}=4.5$ Mm, 2fl$\alpha$ at $t=591.5$ s, the last time shown in the right column, bottom row of Figure \ref{fig:overview}.
In the panels of the neutral and charged densities (top panels) the orange arrows show the velocity of neutral and charged fluid, respectively.
We overplot the magnetic field lines and the decoupling velocities over the current density map (bottom-left panel) and isocontours of the
temperature decoupling ($T_c-T_n$) { in the range $-1.8\times10^6$~K and $1.8\times10^6$~K} over the charged temperature image (bottom right panel). 
}
\label{fig:z45bottom}
\end{figure*}

\subsection{Analysis for coronal reconnection}

We next study the case $z_{\rm rec}=4.5$~Mm where the reconnection was triggered in the coronal region. A zoomed view shown in Figure  \ref{fig:z45bottom} shows what happens near and above the TR, after the downwards reconnection outflow hits 
the dense material in the chromosphere and is reflected. 
The snapshots of density of neutrals versus charges (top row) show again a pattern of evacuated versus enhanced regions with dense versus evacuated edges. 
The decoupling in velocity (bottom left panel of Figure~\ref{fig:z45bottom}) points across the magnetic field lines and the isocontours of the decoupling in temperature ($T_c-T_n$), 
{ shown in the bottom right panel,} follow the  structures in current density,
suggesting that the decoupling is related to the magnetic fields.
Smaller current sheets are formed at different locations and with different orientation, 
meaning that this separation of neutral versus charged densities across the current sheet is generally related to reconnection and not determined by gravity or localized resistivity.
In a time evolution (an animation z4.5-2flalpha.mp4 is provided) we see that this process of separation seems to reverse in this case, when the structure rises again, but this is influenced by mixing with both neutral and charged material coming from below.
Hence, in both 2fl$\alpha$ cases, we find clear evidence of a nonlinear runaway process that enhances the spatial separation between charged and neutral fluids near current sheets.
This runaway process is clearly related to having a large collisional free path, as compared to the width of the CS, as it is not seen at all in the 2fl model.

Overall, we identified a runaway (decoupling) instability in a weakly collisional regime, which occurs in a non-stationary two-fluid setup to the reconnection problem. This will ultimately break down the fluid assumptions if neutral and charged densities decrease towards zero in separate locations.
The exact conditions for the onset of this nonlinear instability, such as the $\alpha$ collisional parameter range, can be subject of a more idealized (without gravity) setup in the future.

\subsection{Reconnection rate}
\begin{figure}
\FIG{
\includegraphics[width=8.5cm]{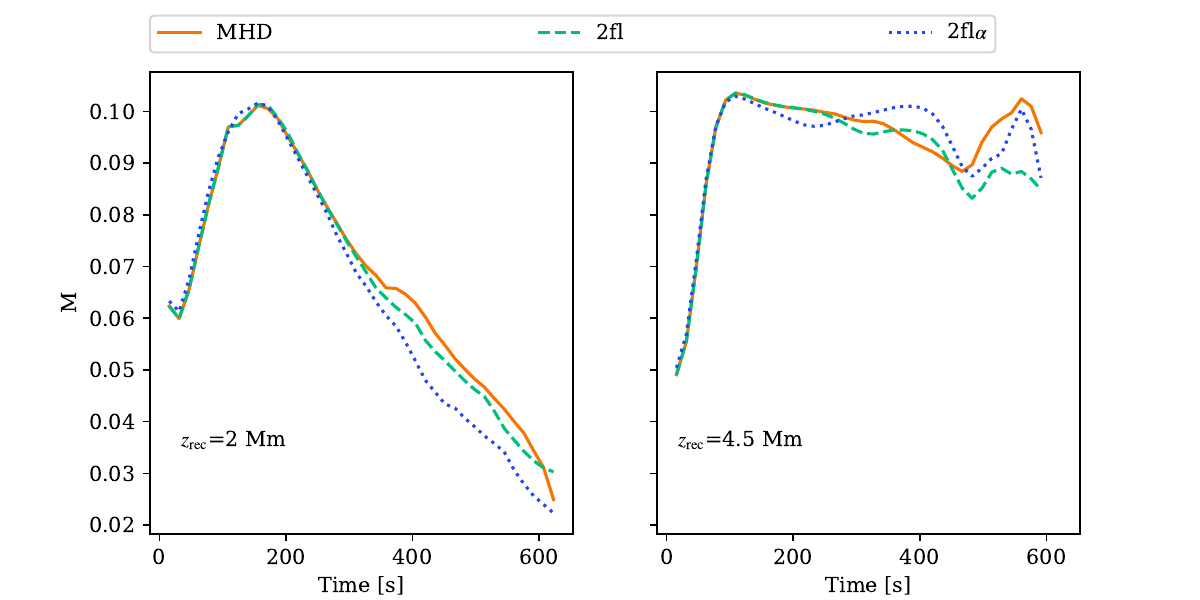}
}
\caption{
Reconnection rates for the six simulations. For $z_{\rm rec}=2$ Mm (left panel) and $z_{\rm rec}=4.5$ Mm (right panel) as a function of time. Different curves indicate different models:
MHD (orange solid line), 2fl (green dashed line) and 2fl$\alpha$ (blue dotted line).
}
\label{fig:rrTime}
\end{figure}

We then calculate the reconnection rate
\begin{equation}
M=\frac{\eta^{X} J_y^{X}}{v_A^{*} B_z^{\rm up}}\,,
\end{equation}
in the same way as \cite{slavaRec2,murtas}. 
The superscripts indicate the points where quantities are evaluated.
The reconnection point $X$ is defined as the point where the dissipation $\eta J_y^2$ is maximum (we included $\eta$ in this calculation because of the localized resistivity).
$X$ will be located at the center of the current sheet ($x=0$) at some height, which might not be the initial reconnection point $z_{\rm rec}$, because this reconnection point migrates vertically,
especially in the TR reconnection case $z_{\rm rec}=2$ Mm, where the stratification is stronger.
{ Initially, the $X$-point is located at the height $z=z_{\rm rec}$, which is one of the parameters of the simulations.}
The border of the CS, indicated by the superscript $^{\rm up}$, is defined as the point located at the same height as $X$, but displaced along the $x$-direction, as it is located at the point where the current density is half
the current density measured at $X$. Because of the horizontal symmetry, the values calculated at the two opposing symmetric points at the borders of the CS are averaged, because numerically exact left-right symmetry might not be preserved. 
The Alfv\'en velocity $v_A^*$ is calculated using the { total} density evaluated at the reconnection point $X$ and the magnetic field $B_z^{\rm up}$.
We plot the thus computed reconnection rate as a function of time in Figure~\ref{fig:rrTime}, for both TR and coronal reconnection, each time comparing MHD and both two-fluid rates.
The reconnection rate has an initial increase, which is steeper for $z_{\rm rec}=4.5$~Mm than $z_{\rm rec}=2.0$~Mm.
This is because at $z_{\rm rec}=4.5$~Mm the density is lower and the CS thins faster, however the minimum width of the CS achieved during the simulations
is similar for both heights. 
The maximum reconnection rate reached after this increase phase is $M\approx 0.1$, a typical value for reconnection scenarios which use localized resistivity.
Then, it decreases for both cases, mainly because  the magnetic field dissipates. 
The reconnection rate decreases much faster in the  TR reconnection case that started at $z_{\rm rec}=2.0$~Mm, because of the quick and continuous upwards migration of the X-point over a distance of $\approx$0.5~Mm at the end of the simulation.  It is seen that the reconnection rates remain similar for the MHD and the two two-fluid setups. 
{ It is known that the plasmoid formation increases the reconnection rate \citep{slavaRec1,murtas}, however, in our case the plasmoids appear too late in the simulations to influence significantly the growth rate. }

\subsection{Secondary plasmoids}\label{sec:plasmoid}

\begin{figure}
\FIG{
\includegraphics[width=8.5cm]{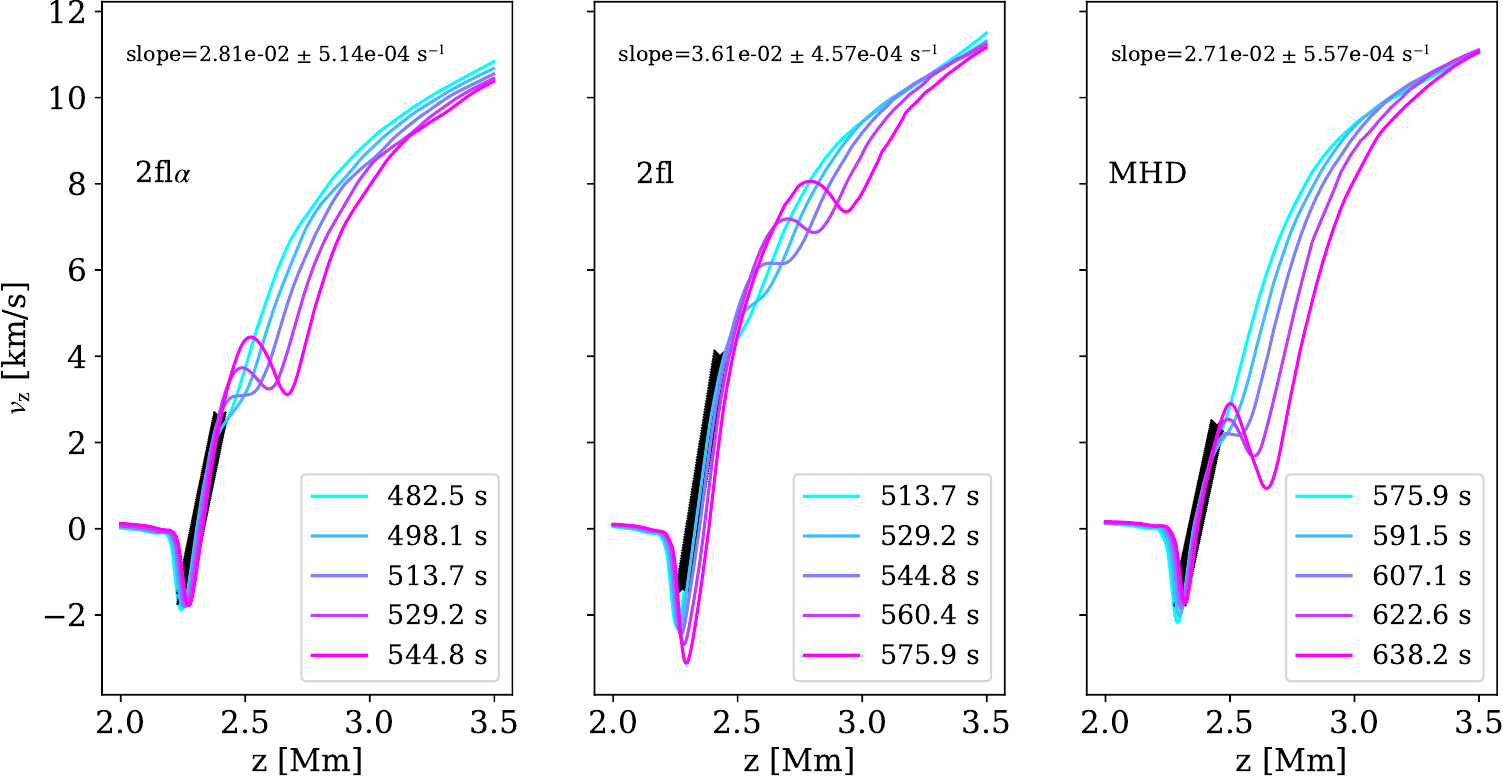}
}
\caption{
Vertical profiles of the vertical velocity when the plasmoids form for $z_{\rm rec}=2$~Mm. Shown for five moments of time in the late stage of the simulations, for the three models: 2fl$\alpha$ (left panel), 2fl (middle panel) and MHD (right panel).
For the two-fluid models (left and middle panels) we show the velocity of charges. The linear fit of the velocity profile around the reconnection point (where the velocity changes sign)
is shown by a black line and the value of the slope is indicated at the top of each panel.
}
\label{fig:plasmo_vel}
\end{figure}
In the last snapshots of the simulations with TR reconnection ($z_{\rm rec}=2$~Mm) we can observe the formation of secondary plasmoids, but not for the coronal case $z_{\rm rec}=4.5$~Mm.
{ Their shape can be clearly seen in the current density map in Figure~\ref{fig:compjy} for the 2fl$\alpha$ case, where we can visually estimate a size of $\approx$0.3~Mm.}
The initial formation phase of the plasmoids can be best seen in the vertical profile of the vertical velocity.
Several equidistant moments of time are shown in Figure \ref{fig:plasmo_vel} for the three models: 2fl$\alpha$, 2fl and MHD. In these profiles, the plasmoids are seen as secondary velocity extrema that develop and get advected upwards. 
The plasmoids move with the outflow from the primary reconnection point, which are at the locations where the vertical velocity changes sign (fitted locally by linear slope in the three panels). We can visually estimate that the fastest growing length scales are similar
in the three cases, however the growth is largest in the MHD case, followed by 2fl$\alpha$ and then 2fl. 
This suggests that the two-fluid effects do not affect the initial phase of the growth of these plasmoids. We do expect that the later stages of this plasmoid evolution will be similarly affected by the runaway decoupling process in the 2fl$\alpha$ case, when charges accumulate towards the middle of the CS.

The linear growth of the tearing mode is affected by the gradient in the vertical flow (parallel to the CS), as shown in a linear resistive MHD analysis about a 2D configuration by 
\cite{stabOutflow}. In their analysis the vertical profile is assumed linear $v_z(z)=a z$, where $a$ quantifies the vertical velocity variation. 
\cite{stabOutflow} show that the growth rate $\gamma(a)$ of the tearing mode at a finite $a$, is reduced from the case without flow gradient, $a=0$, in the sense that $\gamma(a) \approx \gamma(0)-a$. 

A linear analysis of the tearing mode in the two-fluid approach, using simplified assumptions of uniform density, and hence no gravity (and no added vertical flow) is
presented in Appendix~\ref{app1}. This
shows that two-fluid effects will define an effective density $\rho_{\rm c0} \le \rho_{\rm eff} \le \rho_{\rm T0}$ between the background charged $\rho_{\rm c0}$ and total $\rho_{\rm T0}$ densities.
The linear growth of the tearing mode in this simplified two-fluid assumption is bounded between the growth calculated in the MHD assumption when the density is the total density ($\rho_{\rm T0}$) and the charged density ($\rho_{\rm c0}$).  In our setup the ionization fraction is high, and the difference
that the two-fluid effects might introduce in the growth of the tearing mode are of the order of $10^{-4}$~s$^{-1}$. This is shown in our appendix, and can be seen as the difference on the $y$-axis of Fig.~\ref{fig:gr2f} between the orange point which corresponds to the collisional parameter used in 2fl$\alpha$ simulations and the red point which correspond to a value of the collisional parameter larger than the maximum value of $\alpha$ in 2fl cases. 
This two-fluid related difference is
one order of magnitude smaller than the obtained difference in the velocity slope, which we show for the three cases in Figure \ref{fig:plasmo_vel}.
The ordering of the growth rates of the plasmoids, as visually estimated from Figure~\ref{fig:plasmo_vel}, is reversed compared to the ordering of these slopes. This is consistent with the expected reduction in tearing growth rates due to added vertical velocity gradients. Similarly, the large (vertical) gradients in the outflow velocities are probably the reason why we do not observe secondary plasmoids in the simulations
with $z_{\rm rec}=4.5$~Mm, since then the vertical velocities and corresponding vertical gradients are larger than for the $z_{\rm rec}=2$~Mm case.
{ Therefore, in our simulations, the initial linear growth of the plasmoids is rather influenced by the fact that they form at different moments and slightly different heights
during the simulations, when the vertical gradients in the velocity are different. The difference in the growth rate produced by this gradient in velocity is much larger  than the difference in the growth introduced by the collisional effects.}

\begin{figure}
\FIG{
\includegraphics[width=\columnwidth]{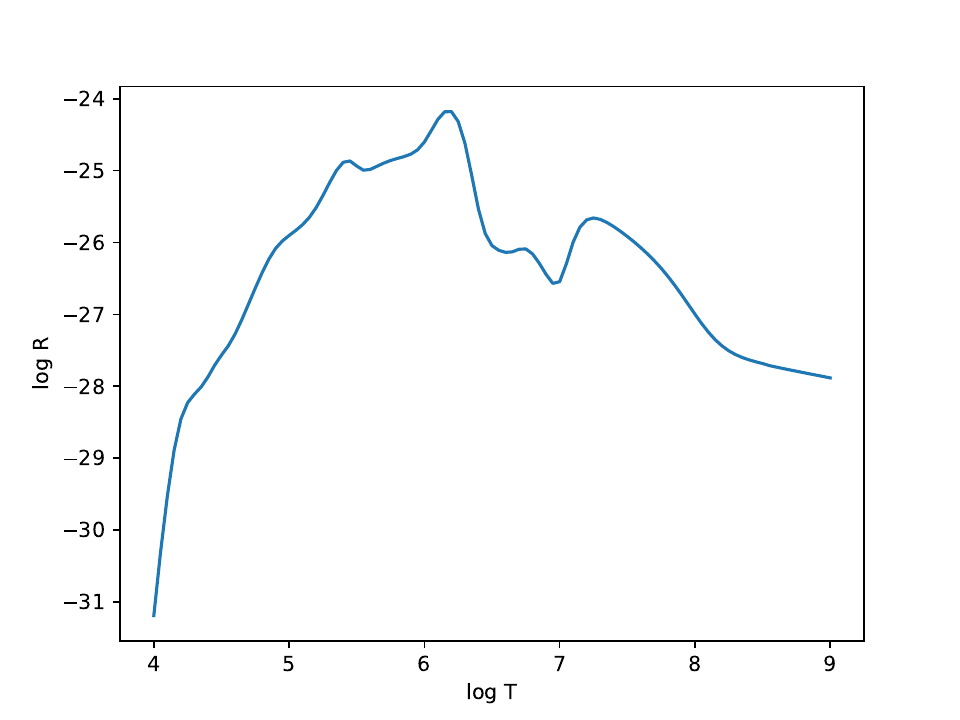}
}
\caption{The 
AIA 193 \AA $ $  response function $R(T)$ as a function of temperature (measured in K) \citep{chianti}. Both axis are logarithmic.}
\label{fig-aiaresp}
\end{figure}

\begin{figure}
\FIG{
\includegraphics[width=\columnwidth,height=6cm]{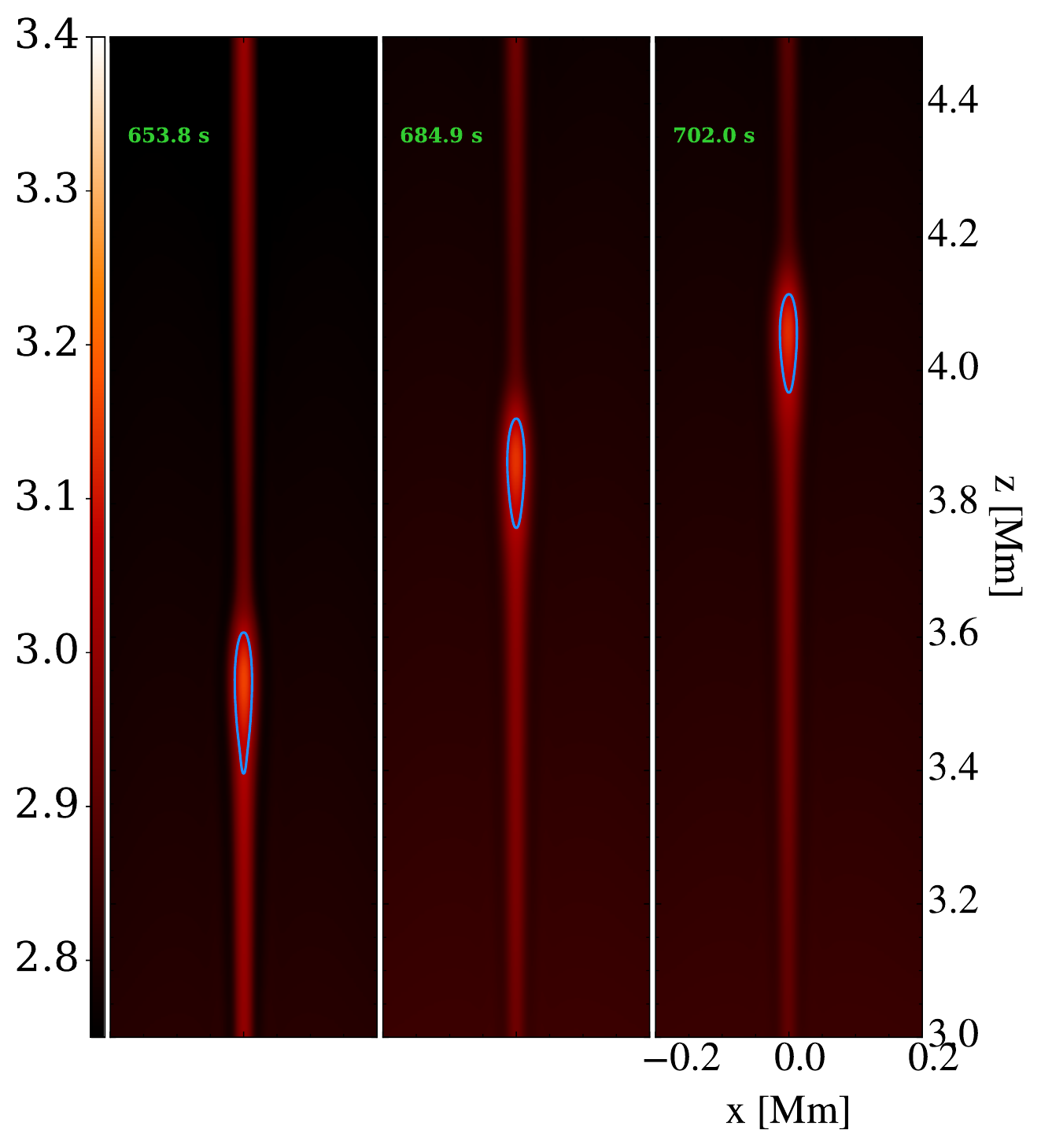}

\includegraphics[width=\columnwidth,height=6cm]{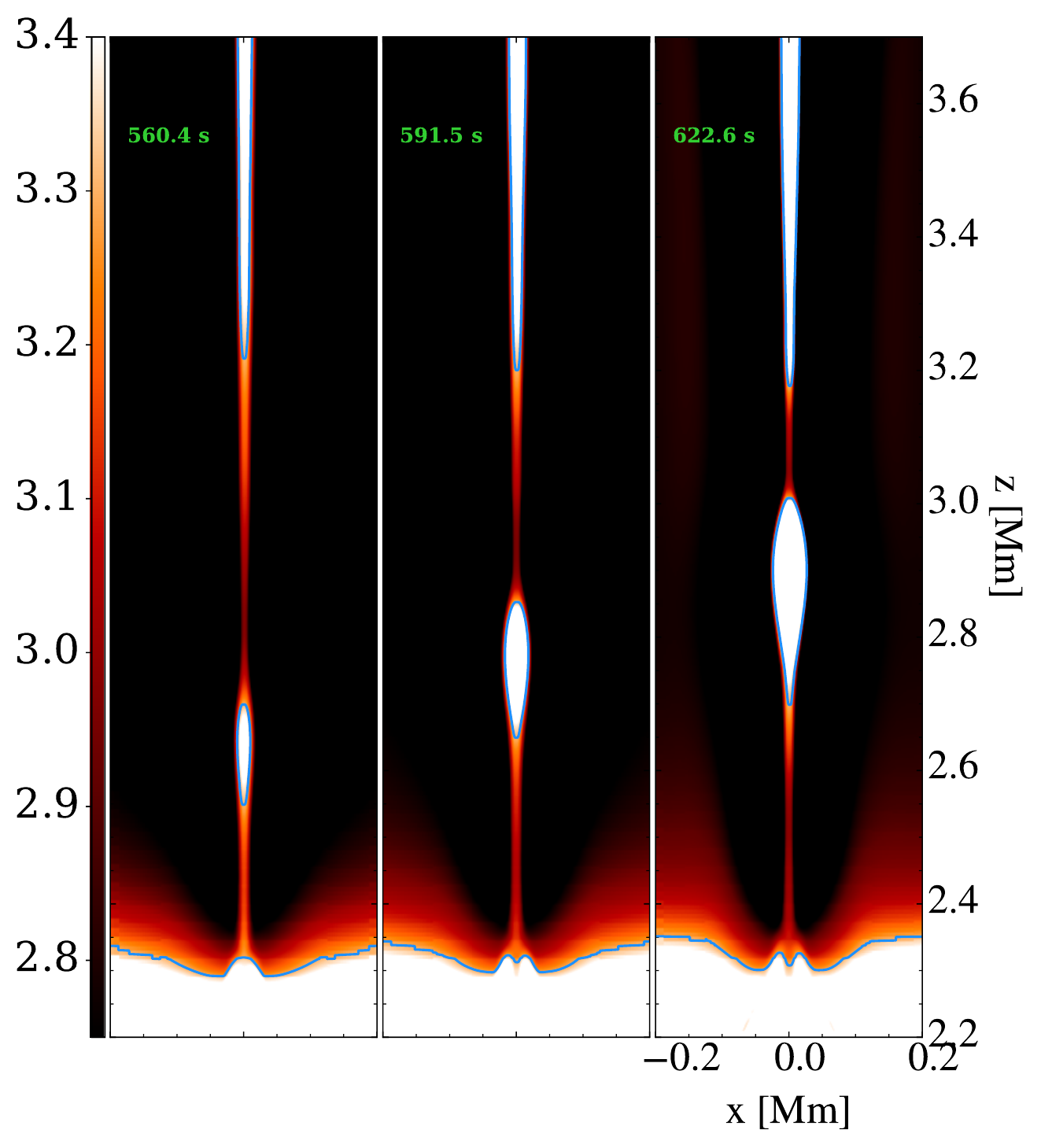}

\includegraphics[width=\columnwidth,height=6cm]{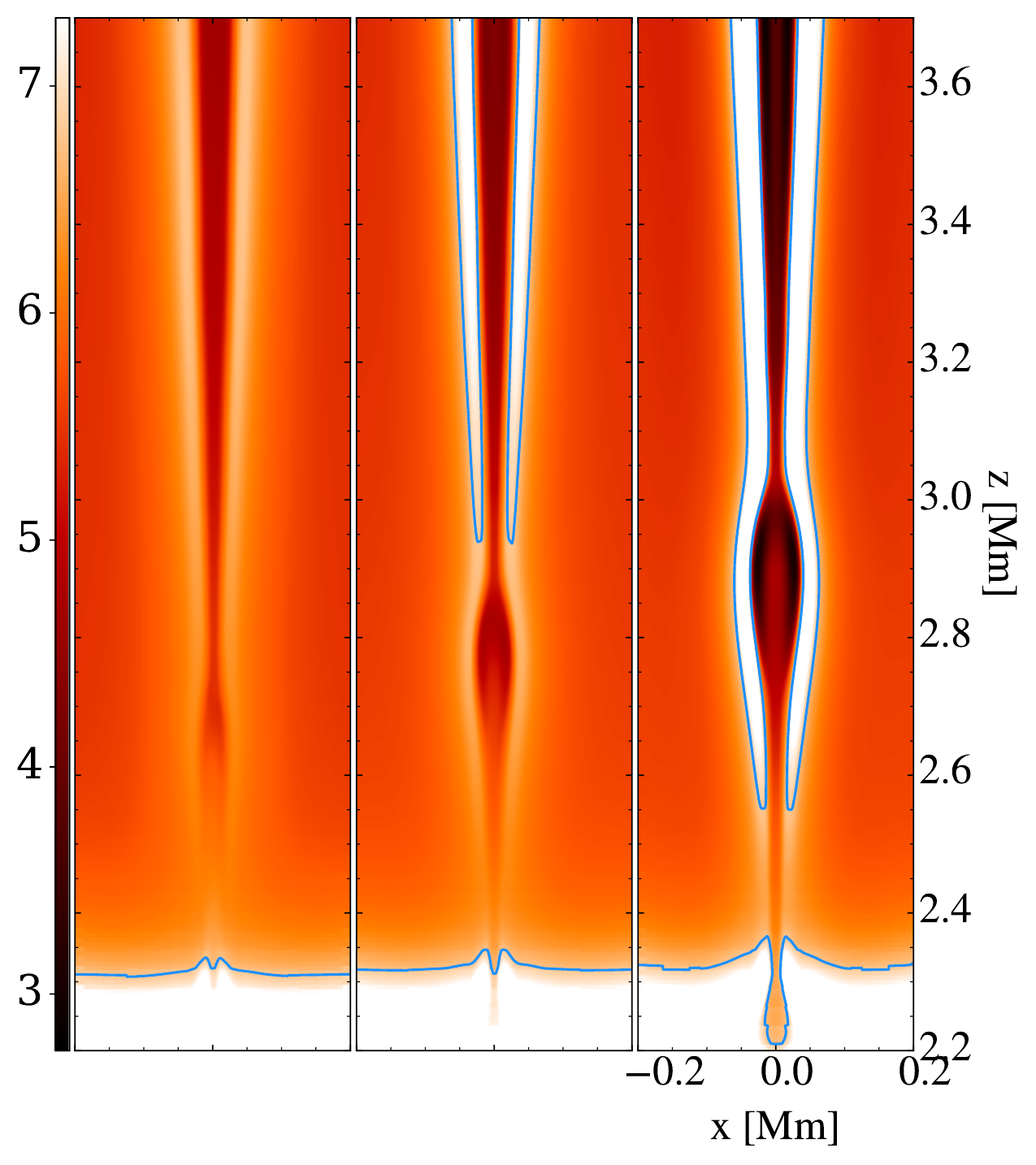}

}
\caption{Synthetic views on secondary plasmoids, at three consecutive times. For plasmoids in the case $z_{\rm rec}=2$~Mm, we show 
synthetic images of AIA 193\AA $ $  emission. 
Top row: 2fl case; middle and bottom rows: 2fl$\alpha$ case.
For the calculation of emission, we used the density and temperature of charges in the top and middle row,
and the total density and the center of mass temperature of neutrals and charges for the bottom row.
The units are non-dimensional. 
We kept the colorbar in the figures, for comparison purposes, as normalization is done in the same way.
The last row shows different limits on the colorbar because of the increased density when total density is considered instead of the charged particles only.
}
\label{fig:plasmo}
\end{figure}

\section{Synthetic views}
\label{sec:synv}

As a relatively straightforward observational validation, we can produce synthetic images resulting from emission from optically thin spectral lines. To do so, we calculate the emission in the Solar Dynamics Observatory Atmospheric Imaging Assembly (SDO/AIA) channel AIA 193 \AA, which has its highest response for temperatures between $10^6$ and
$2 \times 10^6$ K. We do so only above the height $z \approx2$ Mm in our initially stratified atmosphere, because the chromospheric region itself is not appropriately dealt with in an optically thin limit. 
Being optically thin emission, we deal essentially with a function of local temperature $T$ and number density $n$ given by
\begin{equation}
\Lambda(n,T)=n^2 R(T)\,,
\label{eq:emThin}
\end{equation}
where $R(T)$ is a response function which depends on temperature, and a $\log R-\log T$ view is shown in Figure \ref{fig-aiaresp} for the wavelength 193~\AA, obtained using the CHIANTI atomic database \citep{chianti}.
Images in these Extreme Ultra-Violet (EUV) channels of AIA can be synthesized from 3D data cubes by integrating
Eq.~\ref{eq:emThin} along the line of sight (LOS). We have 2.5D data, with an assumed invariance in the third ($y$) direction. As we have 2.5~D simulations, the images are shown at the simulation resolution, and there is no integration along LOS, also making units meaningless. 
As the AIA 193 channel is emitted by Fe XII, emission is obviously from charged species. From our two-fluid plasma-neutral model, we must choose which number density and temperature is taken in Eq.~\ref{eq:emThin}, and a more accurate result might be obtained by using only the charged density, instead of the total density.

In particular, plasmoids are usually observed as  bright blobs due to their increased density compared to the surrounding medium.
In this case models which consider the total density in the calculation of the synthetic image (as is usually done from a single-fluid MHD simulation) would give wrong results, opposed to considering the charged particle density.
Figure~\ref{fig:plasmo} shows synthetic images in the AIA 193 channel which capture the temporal evolution of the plasmoids. We found secondary plasmoids when $z_{\rm rec}=2$~Mm, and show synthetic views for the 2fl case (top row) and the 2fl$\alpha$ case (middle row), where we use the 
charged fluid temperature and density, meaning adopt $\Lambda(n_c,T_c)$ in Eq.~\ref{eq:emThin}.
The bottom row shows for the case 2fl$\alpha$ the same snapshots as in the middle row, but instead using the total density and the center of mass temperature defined by Eq. \ref{eq:temp_2fl}, in practice using $\Lambda(n_{T},T^{\rm 2fl})$ in Eq.~\ref{eq:emThin}.

On all panels of Figure~\ref{fig:plasmo}, we show an emission isocontour at a fixed value.
We can observe that the plasmoid fades away as the surface delimited by the isocontour becomes smaller, while it accelerates upwards in the 2fl case (top row).
On the contrary, the plasmoid decelerates and becomes brighter in the 2fl$\alpha$ case (middle row), most probably because of the increasing charged density due to the runaway effect. 
A completely different interpretation results from the images constructed using total density for the 2fl$\alpha$ case (bottom row), where the plasmoids appear as dark structures that get
surrounded by two bright and descending spikes, coming down from $z\approx4$~Mm, where the accumulation of the neutrals outside the CS seems to be maximal (Figure~\ref{fig:compjy}).
Because of the runaway process, the synthetic image which uses total density gives a wrong image, as the neutrals accumulated outside the current sheet (considered in the total density) would not actually contribute to the emission in this line.

\begin{figure}
\FIG{
\includegraphics[width=\columnwidth]{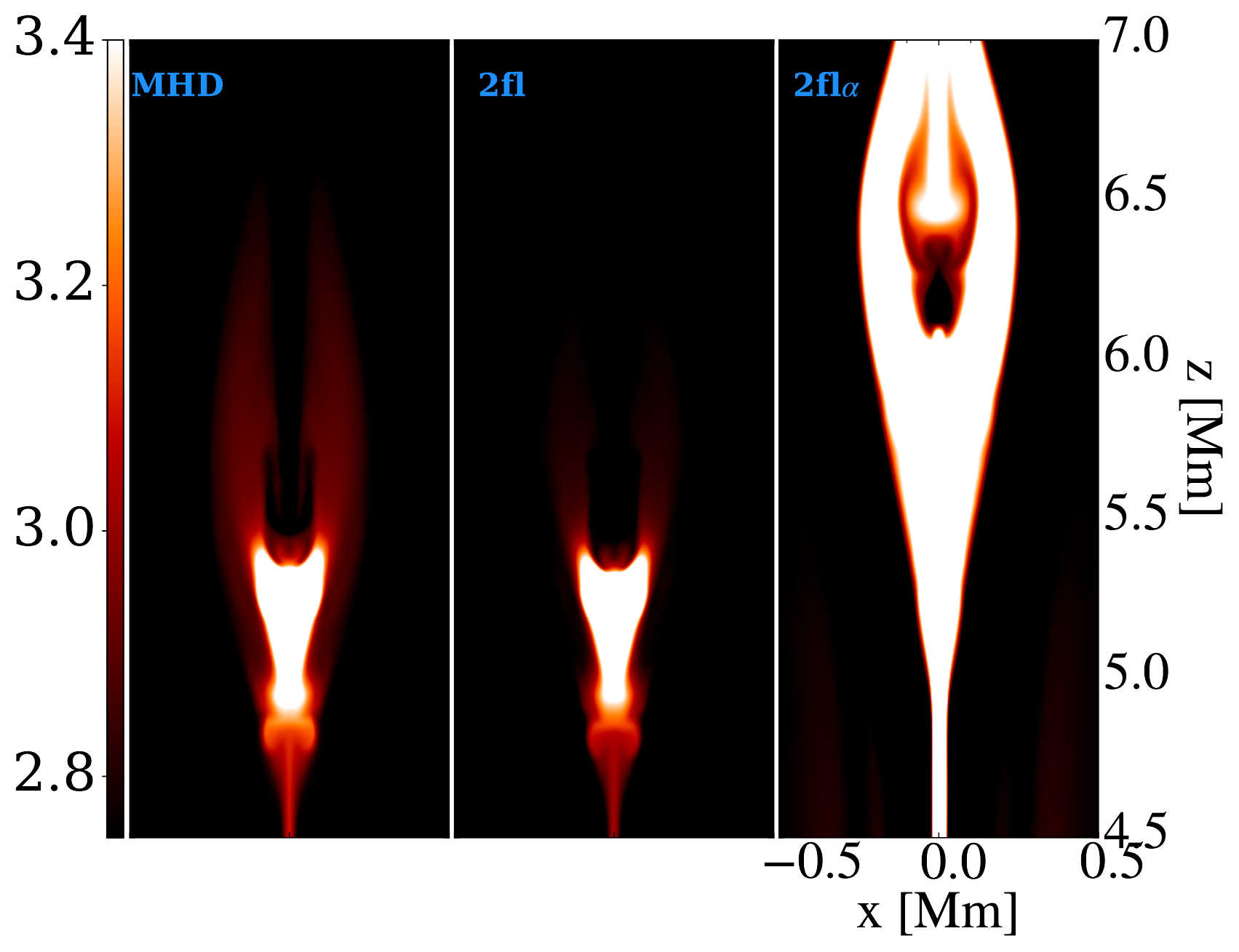}

\includegraphics[width=\columnwidth]{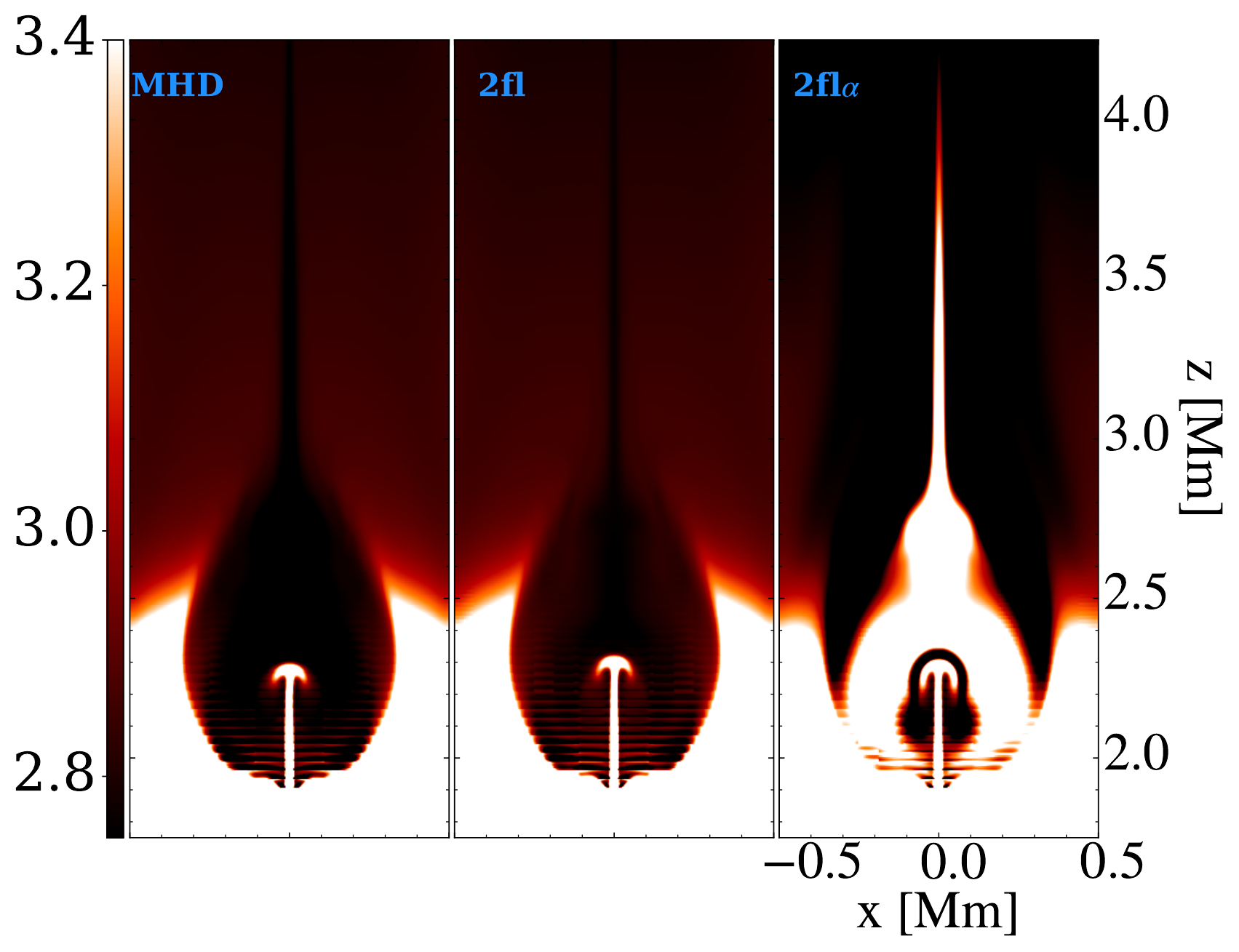}
}
\caption{
AIA 193 \AA $ $  images compared for all six models. Top row: for $z_{\rm rec}=2$ Mm at time $t=622.6$ s. Bottom row: $z_{\rm rec}=4.5$ at time $t=389.1$ s Mm.
The three panels are for MHD (left), 2fl (middle) and 2fl$\alpha$ (right).
Units are non-dimensional.
}
\label{fig-aia1}
\end{figure}

Therefore, charged density only should be used in generating synthetic images in optically thin lines in order to produce  more accurate results.
Finally, we investigate how the MHD model estimates the charged density, and quantify differences obtained in synthetic images for the three models.
The ionization fraction depends on height only and is constant in the MHD model, depending on the background profile. 
The MHD model cannot track how the ionization fraction changes during the simulation because of how the two species evolve, but
we can define a quantity $R$, being the inverse of the normalized mean molecular weight, as
calculated using the two-fluid equilibrium (hence the $0$ subscripts) atmosphere
\begin{equation}
R = \frac{p_{\rm c0} + p_{\rm n0}}{T_0 (\rho_{\rm c0}+\rho_{\rm n0})}\,,
\end{equation}
 which can be used to retrieve the temperature in the MHD model in a consistent way with the two-fluid model
\begin{equation}
T = \frac{p} {R \rho}\,.
\label{eq:temp_mhd}
\end{equation}
As we also know the mean molecular weight of charges and neutrals at each height\footnote{When we use a purely Hydrogen plasma, the normalized mean molecular weights for charges and neutrals are uniform and constant \citep[see Eq.~(12) in ][]{2flpaper},
being equal to 0.5 and 1, respectively.}, then we can
estimate the charged and neutral density from the MHD model:
\begin{equation}
\rho_c = \rho (R-1)\,,\quad \rho_n = \rho (2-R)\,.
\label{eq:rho_cn_mhd}
\end{equation}
With these identifications, we can mimic consistent two-fluid like quantities from a pure MHD model, and then also turn the MHD in a synthetic image based on temperature and charged density.
In practice, for the MHD model the density of charges is then estimated using Eq.~\ref{eq:rho_cn_mhd}.
Figure \ref{fig-aia1} shows the resulting images for the three models and for the two cases of reconnection heights.
We can observe that the MHD and 2fl models give very similar results.
There is a small difference: in the $z_{\rm rec}=2$~Mm the 2fl shows less intensity than the MHD,
but the reverse happens for the $z_{\rm rec}=4.5$~Mm case. This is due to the fact that the density of charges is
slightly overestimated for the fluid coming from below and underestimated for the fluid coming from above. 
Because of different scale heights between charges and neutrals, the ionization fraction is different at different heights.  
This is an intrinsic limitation of the MHD model which only keeps the information of the initial ionization state.
The snapshots used for $z_{\rm rec}=4.5$~Mm are at an earlier time than the final time of this simulation, where we see a clear upward jet feature as the reconnection outflows impact the chromosphere. This jet-like feature with width comparable to the width of the CS ($\approx30$~km) is present in all three images, and may be observable in high cadence, high resolution observations.

\section{Summary}
\label{sec:summary}

We did simulations of two-fluid reconnection in a gravitationally stratified atmosphere, similar to the solar atmosphere,
where we studied the collisional effects for reconnection points situated at different heights.
Because of the localized resitivity used, a Petschek type reconnection developed
with slow shocks propagating along the $x$-direction, disrupting the current sheet in a V shape. The vertical upward velocities of $\approx$ 10~km/s in the $z_{\rm rec}=2$~Mm and the width of the jets, which vary from being similar to the width of the CS ($\approx$ 30~km) near the X-point to $\approx$ 700~km, higher up, due to the slow shocks, are consistent with properties of type I spicules.
The MHD and 2fl simulations showed very similar results.
When $z_{\rm rec}=4.5$~Mm the reconnection outflow hit the denser material below and interacted with reconnected magnetic field,
creating secondary thin current sheets,
leading to locally more turbulent behavior in the post-flare loop region.

The thermal effect of the collisions on the evolution of the neutral fluid has been observed earlier in simplified 1D slow shock simulations, where the heating of the neutrals produces an overshoot in the neutral velocity
\citep{andrewShocks1,andrewShocks2}.
In our case, the decoupling is larger when the collisional effects are increased (the 2fl$\alpha$ cases) and  the heating of the neutrals will produce a runaway effect which separates the neutrals and the charges across the CS. 
The neutrals accumulate outside the CS, while the charges tend towards the center of the CS.
This gives reversed contrasts in the charged density and neutral or total density maps.
The regions with very low density of neutrals have high density of charges (inside the CS) and
the regions with very high density of charges have low density of neutrals (outside the CS), and these differences increase over time.
The temperature maps are consistent with the density maps, with high density regions having low temperatures, while low density has high temperatures. This was analyzed in detail, and a nonlinear decoupling runaway effect was identified.

We obtain high reconnection rates in all the simulations because of the localized resistivity.
The localized resistivity has the effect to bound the current density, similar to an anomalous prescription.
We obtain the maximum reconnection  rate in the  Petschek model, which is 0.1 \citep{vasi}.
Two-fluid effects do not increase our reconnection rates further, as opposed to the results of \cite{slavaRec1,slavaRec2}. Our setup is closer to conditions relevant for the stratified solar atmosphere.

At later time we observe the formation of secondary plasmoids. They are observed for all the models when $z_{\rm rec}=2$~Mm.
The large outflows and associated gradients in the flow when $z_{\rm rec}=4.5$~Mm are likely inhibiting the linear tearing mode. This effect is much more important
than collisions in the early formation of plasmoids. In simplified assumptions of uniform density,
collisions define an effective density in the two-fluid model, similar to the idea presented by \cite{murtas} in simulations of the coalescence instability.
At later stages, however, the accumulation of plasma adds to the nonlinear evolution of the tearing mode.
{ Because of a smaller effective density when the collisional coupling is reduced, the effective Alfv\'en speed is larger and the simulations evolve slightly faster.}

The secondary plasmoid formation process would look completely different in synthetic images which use total density instead of charged density in the calculation
of emission in optically thin lines.
When the charged density is used for the synthetic image generation, an MHD model which does not keep track of the ionization fraction produces slightly different results because of this limitation. It became evident that using charged particle densities leads to more realistic behavior, in line with observed enhanced emission blobs.  

\begin{appendix}
\section{Linear tearing in two-fluid settings}
\label{app1}

The linearized incompressible, resistive MHD equations, where the background density $\rho_0$ is uniform, the background magnetic field
$\mathbf{B_0}=\left(0,B_{\rm y0}(x),B_{\rm z0}(x) \right)$
 is force-free, and the evolution of the perturbed magnetic field $\mathbf{B_1}$ neglects the 
changes due to the Ohmic diffusion of the equilibrium magnetic field, are:
\begin{eqnarray}
\label{eq:eq1}
\rho_0 \frac{\partial \mathbf{v}}{\partial t}  = \mathbf{J_0}\times\mathbf{B_1} + \mathbf{J_1}\times\mathbf{B_0}\,,  \\
\label{eq:eq2}
\frac{\partial \mathbf{B_1}}{\partial t}  = -\mathbf{\nabla} \times (\mathbf{v} \times \mathbf{B_0}) +  \eta_0 \nabla^2\mathbf{B_1}\,,  \\
\label{eq:eq3}
\mathbf{\nabla}\cdot\mathbf{v}=0\,.
\end{eqnarray}
In these equations, $\mathbf{J_1}=\nabla\times\mathbf{B_1}$ is the perturbed current.
 In a 2.5D geometry ($xz$ plane) we consider the $y$-component after taking $\mathbf{\nabla}\times$ of Eq.~\ref{eq:eq1}, the $x$-component of  Eq.~\ref{eq:eq2}, and
\begin{equation}
\mathbf{\nabla} \cdot \mathbf{B}=0\,,
\end{equation}
which is equivalent in the linear assumption to the  $z$-component of  Eq.~\ref{eq:eq2}.
After assuming a solution of the form $\{v_x(x,z,t), B_{\rm x1}(x,z,t)\} = \{u_x(x),b_x(x)\}\cdot \text{exp}(\gamma t - i k_z z)$, the following system is obtained \citep[see also][]{furth,velli,loureiro}:

\begin{align}
\label{eq:eqlin1}
& i \gamma  \rho_0 \left(\frac{d^2 u_x}{d x^2} -  k_z^2 u_x\right) = \nonumber\\ 
& \quad\quad\quad k_z \left[ B_{\rm z0} \frac{d^2 b_x}{d x^2} - b_x \left( k_z^2  B_{\rm z0} + \frac{d^2  B_{\rm z0}}{d x^2}\right) \right]\,,
 \\
\label{eq:eqlin2}
& \gamma b_x = -i k_z B_{\rm z0} u_x + \eta_0 \left( \frac{d^2 b_x}{d x^2} - k_z^2 b_x \right)\,.
\end{align}
In the two-fluid model, when there are separate momentum equations for charges and neutrals, Eqs.~\ref{eq:eq1}-~\ref{eq:eq3} are replaced by:
\begin{eqnarray}
\rho_{\rm c0} \frac{\partial \mathbf{v}}{\partial t}  = \mathbf{J_0}\times\mathbf{B_1} + \mathbf{J_1}\times\mathbf{B_0} + \alpha \rho_{\rm c0} \rho_{\rm n0} \left( \mathbf{v_n} - \mathbf{v_c} \right)\,,  \\
\frac{\partial \mathbf{B_1}}{\partial t}  = -\mathbf{\nabla} \times (\mathbf{v_c} \times \mathbf{B_0}) +  \eta_0 \nabla^2\mathbf{B_1}\,,  \\
\mathbf{\nabla}\cdot\mathbf{v_c}=0\,,\\
\rho_{\rm n0} \frac{\partial \mathbf{v_n}}{\partial t}  = \alpha \rho_{\rm c0} \rho_{\rm n0} \left( \mathbf{v_c} - \mathbf{v_n} \right)\,.
\end{eqnarray}
In the two-fluid assumption Eq.~\ref{eq:eqlin2} remains unmodified, and Eq.~\ref{eq:eqlin1} can be rewritten, so we now get the following coupled system of two governing linear ordinary differential equations:
\begin{align}
\label{eq:eq2flin1}
& i \gamma   \rho(\gamma)  \left(\frac{d^2 u_x}{d x^2} -  k_z^2 u_{\rm cx}\right) = \nonumber\\ 
& \quad\quad\quad  k_z \left[ B_{\rm z0} \frac{d^2 b_x}{d x^2} - b_x \left( k_z^2  B_{\rm z0} + \frac{d^2  B_{\rm z0}}{d x^2}\right) \right]\,,\\
\label{eq:eq2flin2}
& \gamma b_x = -i k_z B_{\rm z0} u_{\rm cx} + \eta_0 \left( \frac{d^2 b_x}{d x^2} - k_z^2 b_x \right)\,.
\end{align}
where 
\begin{equation}
\rho(\gamma) = \rho_{\rm c0} \frac{\gamma + \alpha \rho_{\rm T0}}{\gamma + \alpha \rho_{\rm c0}}\,, 
\label{eq:rhoeff}
\end{equation}
and $\rho_{\rm T0} = \rho_{\rm c0} + \rho_{\rm n0}$ is the background total density. Since for every $\gamma$ we have $
\rho_{\rm c0}\le \rho(\gamma) \le \rho_{\rm T0}$,
this defines an effective density in the two-fluid model, based on the collisional coupling $\alpha$.

\begin{figure}
\FIG{
\includegraphics[width=8.5cm]{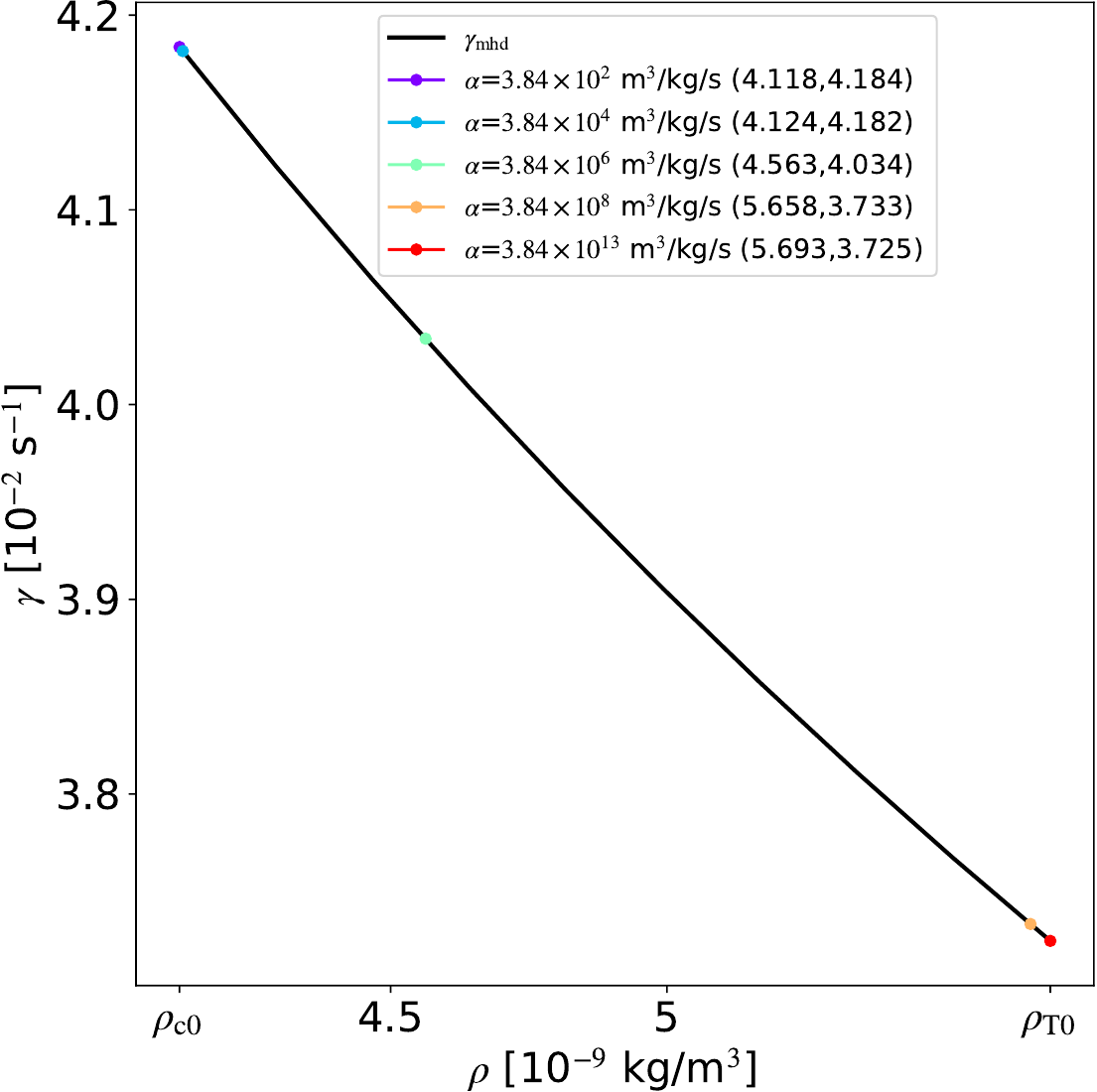}
}
\caption{Growth rate and effective density for the tearing mode with wavelength = 0.4~Mm, the fastest growing mode for $\eta_0=8$~$\Omega m$ and densities at $z=3.5$ Mm. The growth rate obtained in the MHD approximation is shown by a solid black line.
The growth rates in the two-fluid approach are located on this line, depending on the value of the collisional parameter $\alpha$.
Several values of $\alpha$ are considered, as indicated in the legend, where we also show in the brackets the $(\rho,\gamma)$ coordinates of the points.
 }
\label{fig:gr2f}
\end{figure}

We observed in Fig.~\ref{fig:plasmo_vel} that all plasmoids formed have a similar length scale, estimated as $\lambda \approx 0.3$~Mm. 
We consider the densities at $z=3.5$~Mm, the value of the resistivity $\eta_0=8$~$\Omega m$  and solve numerically (using python numpy solver) the eigenvalue problem resulted from the spatial discretization of Eqs.(\ref{eq:eqlin1}),(\ref{eq:eqlin2}), with boundary conditions where we set to zero both variables, thus obtaining the growth rate in the MHD assumption.
Doing so, we found that the largest growing mode (with wavelength larger than 0.1~Mm and
smaller than 0.5~Mm) has wavelength $\lambda=0.4$~Mm for the parameters considered. 
The density gradient scale height at $z=3.5$~Mm is much larger than the vertical size of the domain and than the wavelength considered, so this justifies neglecting the gravity and all variations of the quantities in the vertical direction.

Fig.~\ref{fig:gr2f} shows the computed growth rate as a function of density (varying between $
\rho_{\rm c0}\le \rho \le \rho_{\rm T0}$) for the MHD case, and also gives the two-fluid growth rate for several values of $\alpha$ as indicated in the legend. To get the black solid line in Fig.~\ref{fig:gr2f} which quantifies growth rates in the MHD approximation, we fix $\lambda=0.4$~Mm and solve numerically the eigenvalue problem in the MHD approximation for densities with values between the charged density and the total density.
The growth in the two-fluid approach will be the intersection points between the black line and the curve of $\rho^{-1}$, the inverse of $\rho(\gamma)$ from Eq.(\ref{eq:rhoeff}) (only the intersection points are shown in Fig.~\ref{fig:gr2f} and not the curve $\rho^{-1}$).

\end{appendix}

\begin{acknowledgements}
This work was supported by the FWO grant 1232122N and a FWO grant G0B4521N. 
This project has received funding from the European Research Council (ERC) under
the European Union’s Horizon 2020 research and innovation programme (grant
agreement No. 833251 PROMINENT ERC-ADG 2018). This research is further supported by Internal funds KU Leuven, through the project C14/19/089 TRACESpace. 
The resources and services used in this work were provided by the VSC (Flemish Supercomputer Center), funded by the Research Foundation - Flanders (FWO) and the Flemish Government.
\end{acknowledgements}

\bibliographystyle{aa}

\end{document}